\documentclass[useAMS,usenatbib,iop,numberedappendix]{mn2e} 
\usepackage[colorlinks=true,linkcolor=blue,citecolor=blue]{hyperref}
\usepackage{siunitx}
\usepackage{lipsum,graphicx,multicol}
\usepackage{amsmath,amsfonts,amssymb}
\usepackage{makeidx}
\usepackage{graphicx}
\usepackage{epstopdf}
\usepackage{multirow}
\usepackage{color}
\usepackage{tablefootnote}
\usepackage{lscape}
\usepackage{tablefootnote}
\usepackage[T1]{fontenc}
\usepackage{aecompl}
\usepackage[caption=true]{subfig}
\usepackage{adjustbox}
\usepackage{longtable}
\usepackage{supertabular}
\definecolor{ored}{rgb}{1.00,0.27,0.00}
\usepackage{tablefootnote}

\title[Chandra centres for COSMOS groups]{Chandra centres for COSMOS X-ray galaxy groups: Differences in stellar properties between central dominant and offset brightest group galaxies.}
\author[Gozaliasl et al.]{Ghassem Gozaliasl$^{1,2,3}$\thanks{E-mail: ghassem.gozaliasl@utu.fi}, Alexis Finoguenov $^{2,4}$, Masayuki Tanaka$^{7}$, Klaus Dolag$^{12,13}$, \newauthor Francesco Montanari$^{2,3}$,
Charles C. Kirkpatrick$^{2}$, Eleni Vardoulaki$^{6}$, Habib G. Khosroshahi$^{5}$, \newauthor Mara Salvato$^{4}$, Clotilde Laigle$^{18}$,  Henry J. McCracken$^{9}$, Olivier Ilbert$^8$, Nico Cappelluti$^{10}$,\newauthor  Emanuele Daddi$^{14}$,  Guenther Hasinger$^{15}$, Peter Capak$^{16,17}$, Nick Z. Scoville$^{19}$,  Sune Toft$^{29}$, \newauthor Francesca Civano$^{20}$, Richard E. Griffiths$^{21}$, Michael Balogh$^{22}$, Yanxia Li$^{23}$, Jussi Ahoranta$^{2}$,\newauthor Simona Mei$^{24,25,26}$,
Angela Iovino$^{27}$, Bruno M. B. Henriques$^{28}$, Ghazaleh Erfanianfar$^{4}$ \\
 $^1$ Finnish centre for Astronomy with ESO (FINCA), University of Turku, V\"{a}is\"{a}l\"{a}ntie 20, FI-21500 PIIKKI\"{O}, Finland\\  
  $^2$ Department of Physics, University of Helsinki, P. O. Box 64, FI-00014 , Helsinki, Finland\\
  $^3$ Helsinki Institute of Physics, University of Helsinki, P.O. Box 64, FI-00014, Helsinki, Finland\\
$^{*}$The rest of affiliations are listed after the references.}
\begin{document}
\maketitle
\label{firstpage}
\begin{abstract}
We present the results of a search for galaxy clusters and groups in the $\sim2$ square degree of the COSMOS field using all available X-ray observations from the XMM-Newton and Chandra observatories. We reach an X-ray flux limit of $3\times10^{-16}\;ergs\;cm^{-2}\;s^{-1}$ in 0.5--2 keV range, and identify 247 X-ray groups with $M_{200c}=8\times10^{12}-3\times10^{14}\;M_{\odot}$ at a redshift range of $0.08\leq z<1.53$, using the multiband photometric redshift and the master spectroscopic redshift catalogues of the COSMOS. The X-ray centres of groups are determined using high-resolution Chandra imaging. We investigate the relations between the offset of the brightest group galaxies (BGGs) from halo X-ray centre and group properties and compare with predictions from semi-analytic models and hydrodynamical simulations. We find that BGG offset decreases with both increasing halo mass and decreasing redshift with no strong dependence on the X-ray flux and SNR. We show that the BGG offset decreases as a function of increasing magnitude gap with no considerable redshift dependent trend. The stellar mass of BGGs in observations extends over a wider dynamic range compared to model predictions. At $z<0.5$, the central dominant BGGs become more massive than those with large offsets by up to 0.3dex, in agreement with model prediction. The observed and predicted lognormal scatter in the stellar mass of both low- and large-offset BGGs at fixed halo mass is $\sim0.3$dex.
\end{abstract}
\newpage

\section{INTRODUCTION} \label{Introduction}
According to the standard scenario of galaxy formation, galaxies form via 
cooling and condensing gas at the bottom of the potential wells of a population 
of hierarchically merging dark matter haloes \citep{White78}. After a halo and 
its \textquotedblleft central\textquotedblright\; galaxy fall into a larger 
system, it becomes a subhalo and its galaxy becomes a \textquotedblleft 
satellite\textquotedblright. The cold gas of this satellite galaxy  
may be stripped, leading to a sharp decline in star-formation, reddening its 
colour. Strong tidal stripping can eject stars or even disrupt the satellite 
altogether, providing more material for the disc of the central galaxy of the
massive halo and the stars' stellar halo. Consequently, the central galaxy can 
grow, and become the most massive and luminous galaxy in the system 
\citep[e.g.,][]{springle2005cosmological,skibba2010brightest,Guo11,Henriques15}.

Following this paradigm and the $ \Lambda $-dominated cold dark matter ($\Lambda$CDM) model, 
several semi-analytic models have been implemented  using the 
Millennium simulations (I\&II) \cite[e.g.,][]{springle2005cosmological,Bower06,deLucia07,Guo11,Henriques15}.
However, observations show that the \textquotedblleft central galaxy 
paradigm\textquotedblright\; (CGP), which predicts that central galaxies are the most massive and 
brightest cluster/group galaxies, is not always true 
\citep{beers1983environment,sanderson2009locuss,skibba2010brightest}. 
\cite{skibba2010brightest} analysed the offsets of the line-of-sight velocities and projected positions of brightest group galaxies relative to the other group members using the Sloan Digital Sky Survey (SDSS) cluster catalogue of \cite{yang2007galaxy} and ruled out the CGP. 

The assumption of CGP is critical in a number of measurements  such as halo 
mass estimates using satellite kinematics \citep[e.g.,][]{more2008satellite}, strong 
and weak lensing \citep[e.g.,][]{kochanek1995evidence,sheldon2009cross}, halo occupation 
modelling \citep{tinker2008void} and algorithms for identifying groups \citep{yang2007galaxy,Yang09,Yang12}. It is also well-known 
that central galaxies exhibit different characteristics such as size, 
morphology, colour, star formation, radio and active galactic nuclei (AGN) activities compared to the satellite galaxies of the same stellar mass. The dependence of the central galaxy properties on 
the halo properties such as halo mass has been found to be strong \citep{deLucia07,van2008importance,skibba2009halo,gozaliasl2014gap,gozaliasl2016brightest}. 
Admittedly, these results suggest that a precise definition of central galaxies is 
essential for a precise modelling of galaxies and interpreting the 
observational results. This paper investigates the validity of the CGP in X-ray 
galaxy groups quantifying the offset of the projected positions 
of BGGs relative to the peak of the X-ray emissions from the
intragroup hot gas and medium. 

Galaxy evolution is thought to be the result of halo growth, as well as several other galaxy formation processes (e.g.; star formation, feedback from star formation and 
AGN), and environmental effects. To recognize the role of various physical 
processes of galaxy formation and to link  galaxies to their dark matter 
haloes, studies look for the relation between the halo mass 
function and the stellar
mass function. The stellar-to-halo mass (SHM) relation is thought to be related
to the star formation efficiency, and to the strength
of feedback from star formation and AGN. It has broadly been studied as a 
function of time using several techniques such as matching the abundances of 
observed galaxies and simulated dark haloes ranked by stellar and dark matter 
mass \citep{Behroozi10,Moster13}, the conditional luminosity function method proposed by  
\cite{Yang12}, by the halo occupation distribution (HOD) formalism \citep{moster2010constraints,Moster13,Behroozi13}, and by combining the HOD, N-body simulations, galaxy clustering, and galaxy-galaxy lensing techniques \citep{Leauthaud12,Coupon15}.

Observations indicate that there is a strong correlation between the stellar mass of central galaxies and halo mass of hosting haloes, particularly at low halo masses ($ M_{200}\lesssim 10^{12}\; M_\odot $). The stellar mass of satellite galaxies does not show such dependency on halo mass. Both observations and simulations indicate the presence of a large scatter in the stellar mass of central galaxies at fixed 
halo mass \citep{moster2010constraints,Behroozi10,Behroozi13,Coupon15,matthee2016origin}. 

Several studies (e.g.; \citet{Behroozi10,Coupon15,matthee2016origin}) have searched for the 
origin of this scatter and have quantified the different sources of systematic 
errors, such as varying the assumed cosmology, initial mass function (IMF), the 
stellar population model (SPE), and the dust attenuation laws. Despite these 
efforts, the inconsistencies between the observational data and model 
predictions illustrate that scatter in the stellar mass of central galaxies is 
still an unresolved problem. However, the effect of CGP on the scatter of stellar 
mass has not been enunciated yet, while it is well known that the properties of 
galaxies change with increasing the offset between the 
galaxy position and the centre of clusters. The primary goal of this study is 
to address the presence of an offset between the coordinate of the most massive 
 galaxies and the position of the X-ray peak. We construct  the 
stellar mass distribution and compare the corresponding distribution for BGGs 
with low and high-offset from the group X-ray centres. We also examine the 
impact of the offset on the scatter in the stellar mass of the central massive 
galaxies at fixed halo masses.

The COSMOS survey covers $\sim2 $ square-degree  equatorial field and was 
designed to probe the formation and evolution of galaxies, star formation,  AGN and dark matter with 
large-scale structure (LSS) as a function of local galaxy environment and 
redshift out to $z=6$ \citep{scoville2007cosmic}. The COSMOS survey have been observed by a number of 
major space- and ground-based telescopes, notably by the XMM-Newton, Chandra, HST, GALEX, 
MIPS/Spitzer, PACS/Herschel and SPIRE/Herschel, VISTA and SUBARU telescopes, 
and offers a unique combination of deep ($ AB\sim 25-26 $), multi-wavelength 
data ($ 0.25 \mu m \rightarrowtail 24 \mu m $). We use the COSMOS2015 catalogue 
of photometric redshifts of over half a million sources with an excellent  
precision of  ${\sigma }_{{\rm{\Delta }}z/(1+{z}_{s})} = 0.007$
\citep{laigle2016cosmos2015}. The COSMOS field  have frequently been of  the focus of  spectroscopic redshift surveys. The unique data of 
spectroscopic and multi-band photometric redshifts of galaxies  together with the X-ray 
data provided by Chandra 
COSMOS-Legacy Survey \cite{elvis2009chandra,civano2016chandra,marchesi2016chandra}  and XMM-Newton observations allow us to revise the detection 
of X-ray galaxy groups and clusters in COSMOS as previously presented by
\cite{Finoguenov07,George11}. This study aims to improve the determination of 
the position of the X-ray peak (centre) and the redshift of groups and 
clusters. 

This study presents a unique catalogue of 247 X-ray groups of galaxies  
identified in $ 2\;deg^{2} $ of the COSMOS field \citep{scoville2007cosmic} at a redshift range of $ 
0.08\leq z < 1.53$  
with a mass range of $M_{200c}=8\times10^{12}-3\times10^{14} \;M_{\odot}$. High-mass systems in this halo-mass range are on the border line between groups and clusters but for the purpose of this paper we will refer to these systems only as groups. We select the most massive group galaxies within $R_{200}$ 
(where the internal density of haloes is 200 times the critical density of the 
universe). Since the most massive group galaxies are generally the most luminous group galaxies, we will refer to these galaxies as BGGs in this study.  We quantify the projected separation between the position of BGGs and the IGM X-ray emission peaks, defining the BGG offset as the ratio of this angular separation to the group's 
$R_{200}$ and estimate  differences between the stellar properties of the central dominant BGGs and the BGGs with large offsets. We interpret our observational results through a comparison with 
predictions from two semi-analytic models (SAMs) implemented based on the output data of the Millennium simulations by \citet[][hereafter G11]{Guo11} and \citet[][hereafter H15]{henriques2016galaxy}. In addition, for the comparison of our observational results with those from hydrodynamical simulations, we use BGGs and galaxy groups selected from the Magneticum  Pathfinder simulation\footnote{\small www.magneticum.org}, which adopts a WMAP7 \citep{Komatsu11} cosmology (Dolag et al. in prep).

This paper is organized as follows: in section 2, we describe the catalogues of the spectroscopic and photometric redshift of galaxies used in this study. Section 3 describes the procedures for identification of groups, revision of the X-ray centre and redshift of groups, and a description of the new  catalogue of groups. Section 4 presents the sample definition, the BGG selection, the BGG offset from the X-ray centroid, the evolution and distribution of the BGG offset.  It also presents the relations  between the offset with halo mass, the X-ray flux, and the magnitude gap between the first and second ranked brightest group galaxies.
Section 5 presents the differences in the stellar mass of BGGs selected within different aperture sizes: $0.5R_{500}, R_{500}$, and $R_{200}$. It also presents the non-parametric distribution of the stellar mass and the scatter in the stellar mass of BGGs. Section 6 summarizes the results and conclusions. 

Unless stated otherwise, we adopt a cosmological model, with $(\Omega_{\Lambda}, \Omega_{M}, h) = (0.70, 0.3, 0.71$), where the Hubble constant is parametrized as $100 h\; km\; s^{-1}\; Mpc^{-1}$ and
quote uncertainties at the 68\% confidence level.
\section{THE COSMOS SURVEY DATA} 
\subsection{THE COSMOS SURVEY}
The Cosmological Evolution Survey (COSMOS) is a deep multi-band survey centred at $(Ra,Dec)= (+150.1192,+2.2058)$ and covering a 2 $deg^2$ area. The full definition and survey goals can be found in \cite{scoville2007cosmic}.

COSMOS is the largest field that has been observed by  the Hubble Space telescope (HST) so far. In addition, COSMOS guarantees full spectral coverage with multi-wavelength imaging and spectroscopy from X-ray to radio wavelengths by 
the major space-based telescopes (Hubble, Spitzer, GALEX,
XMM, Chandra, Herschel, NuStar) and the large ground-based
observatories (Keck, Subaru, VLA, ESO-VLT, UKIRT, NOAO, CFHT, JCMT, ALMA and others) \footnote{For more information on the COSMOS multi-wavelengths observations,  the list of broad-, intermediate- and narrow-band filters, and the filter transmission that are used by COSMOS, we refer readers to the COSMOS home web page (http://cosmos.astro.caltech.edu/).}

Over  2 million galaxies have been detected in the deep optical images (e.g., i-band) \citep{ilbert2008cosmos}, and 1.2 million in the NIR \citep{laigle2016cosmos2015}, spanning over 2/3 of cosmic time. The Cosmic Assembly Near-infrared Deep Extragalactic Legacy Survey (CANDELS) is also a part of this field which has been surveyed deeper in the NIR with HST \citep{nayyeri2017candels}. The unique multi-wavelengths data set of COSMOS enables a precise determination of the photometric redshift of galaxies \citep[e.g.,][]{laigle2016cosmos2015}. It allows to study the star formation history  and active
galactic nuclei (AGNs) over $z=0.5-6$ \citep[e.g.,][]{karim2011star,ceraj2017cosmic}. Furthermore, the multibands data enables to  detect galaxy groups and clusters \citep{Finoguenov07,George11}, protoclusters, and X-ray group from the core of a high-z protocluster \citep{wang2016discovery}.

\subsection{THE COSMOS SPECTROSCOPIC REDSHIFT SURVEYS}
COSMOS is a unique field in its unparalleled spectroscopic observations. Since 2007, a number of spectroscopic follow-up campaigns have been accomplished in the COSMOS field \citep[e.g.,][]{lilly2007zcosmos,kartaltepe2010multiwavelength,le2015vimos,comparat20150}. The spectroscopic observations of the COSMOS galaxies are still ongoing and  \cite{hasinger2018} present more recently spectroscopic redshifts for 10,718 objects in the COSMOS field, observed through multi-slit spectroscopy with the Deep Imaging Multi-Object Spectrograph (DEIMOS) on the Keck II telescope in the wavelength range $\sim550-980 nm$. The catalogue contains 6617 objects with high-quality spectra (two or more spectral features), and 1798 objects with a single spectroscopic feature confirmed by the photometric redshift.

Table \ref{tab1:specz} provides a list of important characteristics of the spectroscopic redshift surveys. Columns 1 and 2 list the survey name/reference and instrument/telescopes, respectively. Columns 3, 4, and 5 report the number of objects with secure redshift determination, the median redshift, and the redshift range of the survey, respectively. Column 6 show the median i+ band magnitude of galaxies for each survey \citep{laigle2016cosmos2015,hasinger2018}.\\
In this study, we use an updated catalogue of 36274 galaxies with secure spectroscopic redshifts by M. Salvato et al. (2018, in preparation) and \citep{hasinger2018} to determine  the redshift of our groups, when possible.
\begin{table*}
\caption{Characteristics of the spectroscopic redshift samples.  Only the most secure spectroscopic redshifts are considered (those with a flag between 3 and 4). The redshift range, median redshift, and apparent magnitude in the  band are provided for each selected sample.}

\begin{tabular}{llllll}
\hline
Spectroscopic Survey Reference	& Instrument/telescope & $ N_b$ & $z_{med}$ & $z_{range}$ & $i^{+}_{med}$ \\\hline\\
zCOSMOS-bright \citep{lilly2007zcosmos}&	VIMOS/VLT	&8608&	0.48&	[0.02, 1.19]&	21.6\\	
\cite{comparat20150}&	FORS2/VLT&	788	&0.89&	[0.07, 3.65]&	22.6	\\
P. Capak et al. (in preparation),\cite{kartaltepe2010multiwavelength}&	DEIMOS/Keck II	& 2022 &	 0.93& 	[0.02, 5.87]&	23.2\\	
\cite{roseboom2012fmos} &	FMOS/Subaru	&26&	1.21&	[0.82, 1.50]	&22.5	\\	
\cite{onodera2012deep}&	MOIRCS/Subaru	&10	&1.41&	[1.24, 2.09]	&23.9	\\	
FMOS-COSMOS \citep{silverman2015fmos}&	FMOS/Subaru	&178	&1.56 &	[1.34, 1.73]&23.5\\
WFC3-grism \citep{krogager2014spectroscopic} &	WFC3/HST&	11	&2.03	&[1.88, 2.54]	&25.1\\	
zCOSMOS-deep (S. Lilly et al. 2016, in preparation)&	VIMOS/VLT&	767&	2.11	&[1.50, 2.50]&	23.8\\
MOSDEF \citep{kriek2015mosfire}	&MOSFIRE/Keck I	&80&	2.15	&[0.80, 3.71]&	24.2\\	
M. Stockmann et al. (in preparation), \cite{zabl2015emission}	&XSHOOTER/VLT&	14&	2.19	&[1.98, 2.48]	&22.2\\
VUDS \citep{le2015vimos}&	VIMOS/VLT&	998	&2.70&	[0.10, 4.93]&	24.6\\
DEIMOS 10K \citep{hasinger2018} & DEIMOS/Keck II &  6617 & 1 \& 4 & [0.00, 6.00]& 23\\
\hline
\end{tabular}\label{tab1:specz}
\end{table*}
\subsection{THE COSMOS PHOTOMETRIC REDSHIFTS}
When there are not enough galaxies with spectroscopic redshifts within an extended X-ray source to update the redshift of associated group, we revise the redshift of this source and its group using recent photometric redshifts catalogues, notably, the COSMOS2015 catalogue \citep{laigle2016cosmos2015} and the earlier catalogues presented in \cite{ilbert2008cosmos,mccracken2012ultravista,Ilbert13}. All these catalogues use the SED fitting method and  apply the {\small Le Phare code} to measure the photometric redshifts and stellar masses with a  $ \chi^2$ template-fitting method. The details of the method can be found in \cite{ilbert2008cosmos,Ilbert13}.

The COSMOS2015 catalogue contains precise photometric redshifts and stellar masses for over half a million sources. The object detection in this catalogue has been done using  
${{YJHK}}_{{\rm{s}}}$ data from the UltraVISTA-DR2 survey. However for the better estimate of the photometric redshifts, combination of 31 band data has been used. A summary of available data in each band, the average limiting
magnitudes, and the central wavelength of each band  have been presented in Table 1 by \cite{laigle2016cosmos2015}. The COSMOS2015 catalogue is also a unique catalogue in terms of the accuracy of photometric redshifts. Using a secure sample of spectroscopic redshifts such as zCOSMOS-bright 
(see Tab. \ref{tab1:specz}), the precision of the photo-z of galaxies is found to reach  ${\sigma }_{{\rm{\Delta}}z/(1+{z}_{s})}= 0.007$ with a catastrophic failure fraction of $\eta \;=\;0.5$\%. At $3 <z < 6$, the photo-z precision was obtained ${\sigma }_{{\rm{\Delta }}z/(1+{z}_{s})}= 0.021$.  Sec 4.3 and Figure 11 in \cite{laigle2016cosmos2015}  present a detailed analysis on the accuracy of  the photo-z for two type of star-forming and quiescent galaxies with different i-band magnitude ranges from 16 to 27 $mag$. This figure is in agreement with figure 8 in \cite{ilbert2006accurate}, who indicate that the spectral type is not the dominant factor, and that the redshift and the magnitude are more relevant to the photo-z accuracy. We emphasize that early type galaxies interestingly produce a lower quality photo-z (in both \cite{laigle2016cosmos2015,ilbert2006accurate} analysis), probably, because we do not have a sufficiently large variety of templates for this population.

 The COSMOS2015 catalogue covers effective areas of  $0.46\; deg^2$ Ultra deep and $0.92\; deg^2$ of deep UltraVISTA surveys. At the deepest regions, the stellar mass of galaxies reaches a 90\% completeness limit of 
${10}^{10}{M}_{\odot }$ to $z = 4.0$. Details of these regions can be found in section 7.1 (figure 1 and table 7) by \cite{laigle2016cosmos2015}. For more details on the photo-z estimate and the stellar mass estimation, we  refer the reader to \cite{laigle2016cosmos2015}.

For maximizing catalogue completeness for bluer objects and at higher redshifts, \cite{laigle2016cosmos2015} detected objects on a $ \chi^2 $ sum of the ${{YJHK}}_{{\rm{s}}}$ and Subaru  SUPRIME-CAM broad band $z++$ (central wavelength of 910.572 nm) images. However, this catalogue misses around $25\%$ of blue objects that were detected in the i-selected catalogue by \cite{ilbert2008cosmos}. Thus, for a complete identification of groups within the  whole  $\sim2\;deg^2$ area of the COSMOS field and a complete selection of group members, besides the
COSMOS2015 catalogue, we utilize the earlier i-selected v.2 catalogue of photometric redshifts by \cite{ilbert2008cosmos,mccracken2012ultravista}. In addition, \cite{marchesi2016chandra} present a catalogue of 4016 X-ray sources and AGNs in the COSMOS field and measure precise photometric redshift of these objects, we thus use the photometric redshifts of these x-ray sources from  \cite{marchesi2016chandra}. If there is any missing objects and galaxies associated to the extended X-ray emission sources, we determine the overdensity of galaxies using the photometric redshift catalogue presented in \cite{Ilbert13}.  

\section{IDENTIFICATION  OF X-RAY GALAXY GROUPS AND CLUSTERS}
The initial catalogues of the COSMOS X-ray groups were published in \cite{Finoguenov07,George11}. These catalogues combined the available Chandra and XMM-Newton data with developments in the photometric datasets, used for identification of galaxy groups, with confident identification reaching a redshift of 1. They cover mostly massive groups and clusters which are bright in X-rays. For the full details of group identification we refer readers to \cite{Finoguenov07}, \cite{finoguenov2009}, \cite{finoguenov2010x,finoguenov2015ultra}, \cite{George11}, and \cite{Gozaliasl14}.

In this section, we briefly describe the revision of the X-ray centres of the groups using the combined data of Chandra and XMM, application of the red-sequence finder as a primary procedures for cluster  and group identification, and the redshift improvement of galaxy groups relative to their early identification  by \cite{Finoguenov07,George11}. Finally, we assign a quality flag to each group based on a visual inspection of the combined  X-ray 
data of the extended sources and the optical RGB images ($ i,r,g$  broad bands of Suprime-Cam) of galaxies within $ R_{200}$ and present the catalogue.
\begin{figure*}
\includegraphics[width=1\textwidth]{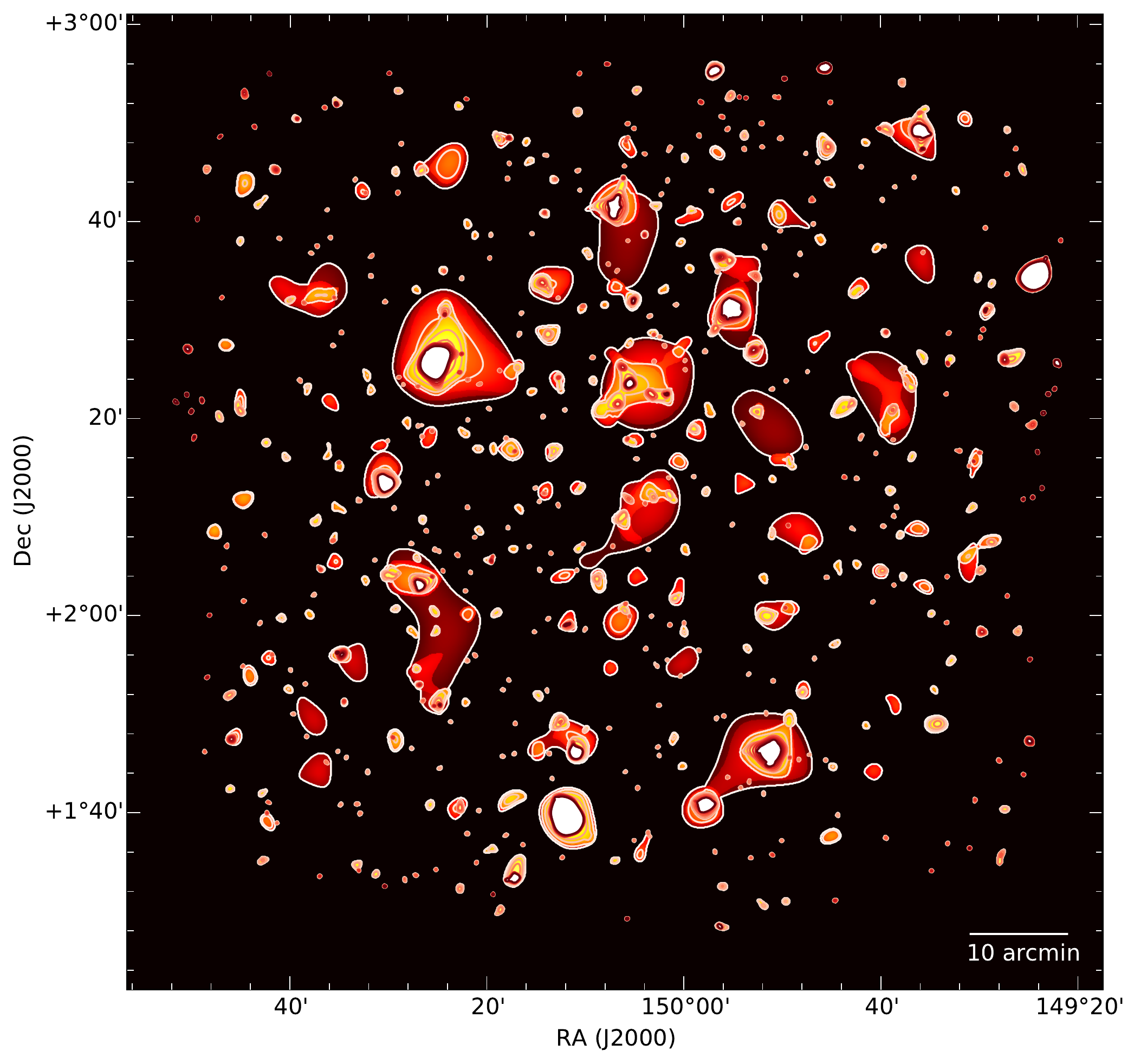}
\caption{The combined Chandra and XMM-Newton $0.5-2\; keV$
wavelet-filtered image of the extended X-ray emission in the COSMOS field. The emission on scales $16-256\; arcsec$ is shown. The white contours denote the level of emission at $6\times10^{-17}$, $3.5\times10^{-16}$ and $1.2\times10^{-15}\;ergs\;s^{-1}\;cm^{-2}\;arcmin^{-2}$ level. The emission on scales of $16\;arc-seconds$ is used to improve the centring. The catalogue of sources corresponds to the signal detected on the $32-128\; arcsec$ scales. The lowest level of the emission corresponds to real detection only for large scale sources with areas of 10 square arc-minutes or more. }\label{xray}
\end{figure*}

\subsection{THE REVISION OF THE GROUP X-RAY CENTRE}
Since then, the visionary Chandra programme has been completed \citep{elvis2009chandra,civano2016chandra}, providing the high-resolution imaging across the full COSMOS field. In addition, the status of photometric data provides robust identification of galaxy groups to a much higher redshift. The revised catalogue of extended X-ray sources in COSMOS, released as a part of this paper, is obtained by co-adding all the existing Chandra and XMM-Newton data in the field. It is very similar to the catalogue used in \citet{George11}, but extends the list of sources beyond the redshift of 1. In addition, we are able to improve on the precision of the centres for extended sources, using the smaller scale emission, detected by Chandra, reducing the statistical uncertainty on the centring from $15^{\prime\prime}$ in \citet{George11} to $5^{\prime\prime}$. The scales of source confusion are also improved from 32 to 16 $arcsec$.  

Following \cite{finoguenov2009}, in this work we consider the detection using the same spatial scales of $32-128\;arcsec$ as employed in our XMM work. On those scales, the combined Chandra data adds $30\%$ to the existing exposure (or $14\%$ in sensitivity), on average, which results in marginal improvements in the catalogue. The main change, possible with Chandra data, is related to the better centring of X-ray emission, as small scales, 16 arc-second scales can also be used. This is a primary importance for the goals of this paper: to separate the BGGs based on the deviation from the X-ray centre. In this work we increase the sensitivity by using combined Chandra+XMM data on 16 arc-second scales after rejecting the possibility of point source contamination using Chandra data on scales of a few arc-seconds, which is sensitive even to three times fainter point sources \citep{civano2016chandra}. 
Figure \ref{xray} shows the combined Chandra and XMM-Newton $0.5-2\;keV$
wavelet-filtered image of the extended X-ray emission sources in the COSMOS field. The emission on scales of $16-256\;arcsec$ is shown. The white contours denote the level of emission at $6\times10^{-17}$, $3.5\times10^{-16}$ and $1.2\times10^{-15}\;ergs\;s^{-1}\;cm^{-2}\;arcmin^{-2}$ level. The emission on scales of $16\;arcsec$ is used to improve the centring. The catalogue of sources corresponds to the signal detected on the $32-128\;arcsec$ scales.
\begin{figure*}
\includegraphics[width=1\textwidth]{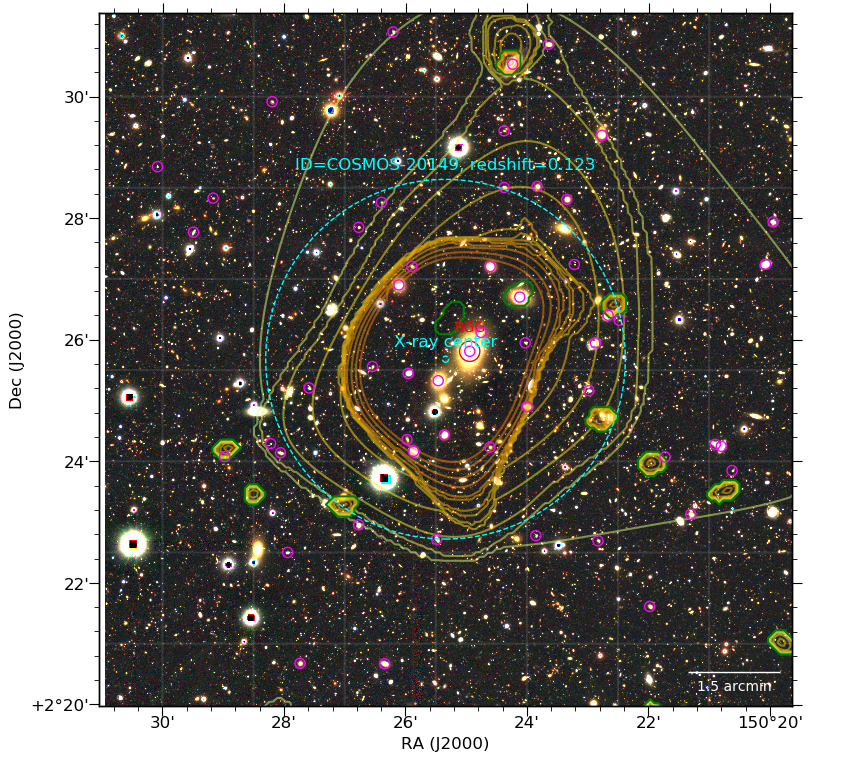}
\caption{The combined RGB images of $i, r$, and $g$ bands with overlaid X-ray emission contours (shown in yellow) for group ID 20149 with 50 spectroscopic members (magenta circles). The BGG (red circle) is close to the centre of X-ray peak (small dashed cyan circle). The large dashed cyan circle shows the group's half $R_{200}$. The small scale X-ray emission sources ($8-16\;arcsec$) are highlighted with green contours. With this scale, we are able to detect the X-ray emission from the bright group galaxies.}\label{get_z}
\end{figure*}
\subsection{THE RED-SEQUENCE APPLICATION}   
In order to ensure the group and cluster identification, we also use the so-called and refined  red sequence method as described in more details in \citet{finoguenov2010x,finoguenov2015ultra}.  This is a further
refinement of the photo-z concentration technique  which is used for identification of groups and assigning their redshift
\citep{Finoguenov07,George11}.

\begin{figure}
\includegraphics[width=0.49\textwidth]{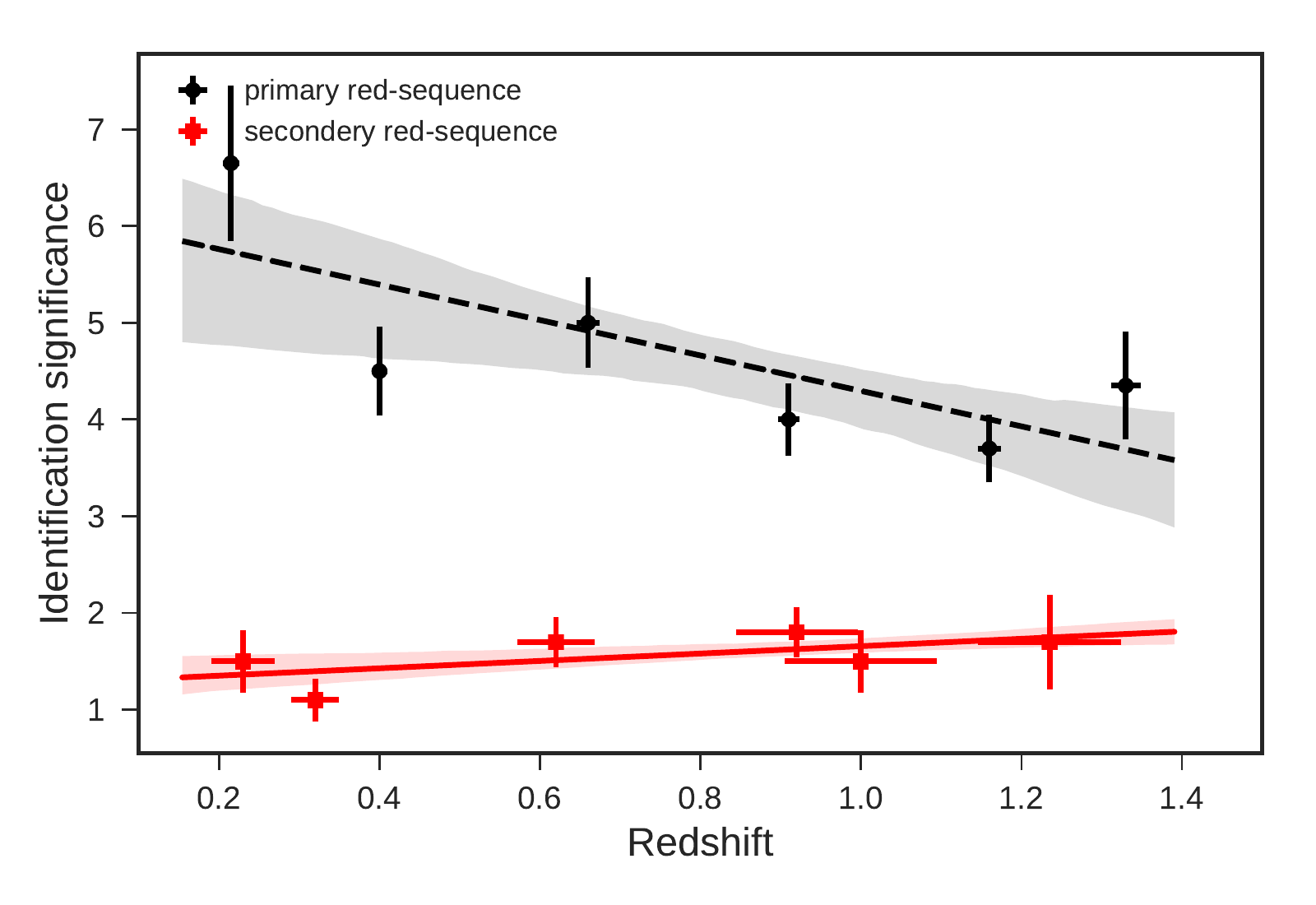}
 \caption{The median of the identification significance versus the redshift of groups when searching for identifying them using the red-sequence method. The black and red data points present the first and the second most significant solutions or the primary and secondary red-sequences corresponding to each extended X-ray source. }\label{signif}
\end{figure}

We run the red-sequence finder for all galaxies located  within each extended 
X-ray emission sources. We apply the red-sequence finder to detect any group candidate at a given redshift within different  aperture sizes from the X-ray centre/peak of each extended X-ray sources. The first  aperture size that we use to select galaxies for the red-sequence test corresponds to $0.5\;Mpc$ (physical) from the centre of
X-ray emission at a given redshift. We also run the red-sequence finder within $R_{500}$ radius of groups which are in common with the \citet{Finoguenov07,George11} catalogues. The application of the red-sequence method can be found in detail in \citet{finoguenov2010x,finoguenov2015ultra}. As described in these papers we measure a redshift for any overdensity of red galaxies at
the position of a group candidate. To quantify the significance of each red-sequence, assume an aperture of the same
size at a random position in the COSMOS field and implement the same procedure
5000 times. We apply a $2\sigma$ clipping when estimating the mean/dispersion of redshift. Thus, group
regions should be clipped out from the mean/dispersion estimates. This provides us with an average number of red galaxies
and its dispersion in the field at a given redshift. The significance of any detected red sequence within an extended X-ray source
is evaluated as a relative overdensity of the group candidate
to that of the field.  Figure \ref{signif} compares the median significance of the first (black data points) and the second (red data points) solutions (red sequences)  versus the redshift. We find that the primary red sequence is always quite significant and more robust than the second red sequence by at least a factor of 2-3 times. The second  red sequence significance is also too low. Figure \ref{signif} also shows that the median significance of the first solution decreases with increasing redshift. We assume the primary red-sequence of each extended source and inspect the  group candidate  by applying the photo-z/spec-z  over-density technique since the red sequence method miss star-forming and  blue group galaxies. 

\subsubsection{THE REDSHIFT DETERMINATION OF GROUPS}
Since time that \cite{Finoguenov07} and \cite{George11} presented the COSMOS X-ray groups catalogues, 
the COSMOS field has frequently been in the focus of multi-band photometric and spectroscopic observations \citep[e.g.,][]{mccracken2012ultravista,le2015vimos,civano2016chandra}. These observations now provide deep, high quality multi-wavelength data ($AB\sim25-26$) in the COSMOS field which cover a wide electromagnetic wavelength range from the X-ray to radio bands. The luxury of 
having a precise multi-band  photometric redshifts \citep[e.g.,][]{laigle2016cosmos2015},  catalogue of  spectroscopic redshifts (M. Salvato et al. in preparation) together with deep X-ray data of Chandra COSMOS-Legacy Survey \citep{civano2016chandra,marchesi2016chandra} motivated us 
to revise the previous catalogue of X-ray galaxy groups in the COSMOS field \citep{Finoguenov07,George11} and also search for new X-ray groups (Gozaliasl et al. in preparation). \\

For groups with spectroscopic data available, we update the photometric redshift with a more accurate spectroscopic redshift using the bi-weight location method described in \citet{beers90}. To avoid the potential contamination due to the presence sub-structure, we consider all group members within an $R_{500}$ aperture and obtain an initial group redshift.  The proper velocity of each member is then computed and a $3\sigma$ clipping is applied to remove any possible projected interlopers.  We iterate over multiple steps until the solution converges. Finally, the redshift is assigned if there are 3 or more spectroscopic members remaining. Groups are visually inspected, especially, where the system are unrelaxed (i.e.; mergers). 

Using the bi-weight location method, we estimate the spectroscopic redshift to groups within $R_{200}$ and  determine its difference with the redshift   estimated using spec-z members within $R_{500}$. Fitting a single Gaussian function to the distribution of $\Delta z(R_{200},R_{500})$  for groups at $0.04<z<1.53$,  $0.04\leq z \leq 0.5$ and $0.5<z\leq1.53$, we quantify the dispersion/standard deviation as: $\sigma_{\Delta z}=0.0038,\;0.0029$, and $0.0045$, respectively.

In Fig. \ref{get_z}, we show the combined  RGB images ($i, r$, and $g$ filters of the Suprime-Cam of Subaru telescope)  with overlaid  X-ray contours for the group ID 20149 at $z=0.123$. This system includes 50 $spec-z$ members within $R_{200}$ (magenta circles). The BGG are marked with solid red circle and X-ray peaks as dashed red circles. 

All groups and  their central galaxies  are visually inspected. We find that 143 out of 247 galaxy groups have at least three members with spectroscopic redshifts within  ($R_{500}$) the redshift of these galaxies matches the photo-z of groups within errors and this allows the precise assignment of spectroscopic redshifts to these groups. When we increase the aperture size to  $R_{200}$, the number of groups with at least three spectroscopic redshift members increases to 183 out of 247. Thus, the redshift of the 40 out of 247 groups is updated considering their spectroscopic redshift members within $R_{200}$. \\
  For the rest of the 64 out of 247 groups that contain less than three spectroscopic members, we assign a photometric redshift using the COSMOS2015 catalogue \citep{laigle2016cosmos2015}, the catalogue of  i-selected sources by \cite{ilbert2008cosmos}, and the catalogue of X-ray sources (e.g., AGNs) by \cite{marchesi2016chandra}. Further, it is noted that 24 groups (from 64 groups) consist of two spec-z members within $R_{200}$ which their redshifts match with the group photo-z within errors. In addition, $188/247$ central group galaxies in our sample are galaxies with spectroscopic redshifts which in turn allows us to examine  further whether the photo-z of groups is accurate. In summary, we find that 27\%  and 20\% of all groups members within $R_{500}$ and $R_{200}$ in our catalogue are spec-z galaxies.\\
 
To revise the redshift of groups with not enough ($<3$) spec-z members, we measure galaxy over-densities in the photometric redshift space similar to the method used by \cite{Finoguenov07} and \cite{George11}.
We select galaxies from the original photometric redshift catalogue which have high quality
photometric redshift determination ($95\%$ confidence interval) and are not morphologically classified as stellar objects.

The precision of photometric redshift allows us to select redshift slices
covering the range $0<z<4$. However, for the current catalogue of groups with the majority of them having large scale X-ray emission, we limit this range to $0.08<z<1.53$. To provide a more refined redshift estimate for the identified structures,
we slide the selection window by a 0.05 step in redshift.
We add each galaxy as one count and apply the filtering techniques presented by \cite{vikhlinin1998catalog} to detect excess in the galaxy number density on scales ranging from $20\;arcsec$ to $3\;arcmin$ on a confidence
interval of $4\sigma$ with respect to the local background. We determine the local
background by both the field galaxies located in the same redshift slice and galaxies contributed to the slice due to a catastrophic failure in the photometric redshift. 

In order to be sure of the measured photometric redshift of a group, we separately use the Kernel density estimation method (KDE) and determine the redshift distribution and probability density functions (PDF) for all galaxies associated with each extended X-ray source within $0.5\;R_{200}$. We then determine the redshifts 
corresponding to the position of the centres of four peaks with high PDF ($>0.4$). We then take these redshifts ($z_{peak}$) and select all galaxies whose redshifts lie at $z_{peak}-z_{err} \leq z \leq z_{peak}+ z_{err}$, where $z_{err}$ corresponds to the $photz$ precision for the given redshift. For each redshift candidate, we measure the significance of each peak and after a visual  inspection, we select the best redshift and update the photometric redshift of the group. 

\subsubsection{THE CATALOGUE DESCRIPTION}
We describe our catalogue of 247 X-ray galaxy clusters and groups identified so far in the COSMOS field. The full  catalogue of galaxy groups is presented electronically. Table \ref{catalogue} lists a sample of these groups with X-ray properties. Column 1 lists the groups and clusters identification ID. The last 3 digits of this ID present the previous identification ID as defined by \cite{Finoguenov07,George11}. If the X-ray centre of groups is defined based on the small and large scale X-ray data, the first digit of the ID begins with 1 and 2, respectively. If the current X-ray centre of previous catalogue needs no correction, the first digit of the ID is 3.  Columns 2 and 3 report the right ascension and declination of the position of the peak of the extended X-ray emission from the intra-cluster/intra-group hot gas  in equinox $J2000.0$. Column 4 presents the redshift of clusters and groups. 

Column 5 lists group's identification flags. We define four quality flags to describe the reliability of the optical and X-ray counterparts as follows. We assign flag 1, if the group has a secure X-ray emission from the IGM and we can define an X-ray centre. In addition, the group has spectroscopic members in which we are able to measure a spectroscopic redshift for the group. Flag 1 groups generally include central BGGs with spectroscopic redshifts. Group 20149 in Fig. \ref{get_z} is an example of a Flag 1 group. Flag 2 shows a group which share the X-ray emission with a foreground/background object and we assign the X-ray flux  between them based on the concentration of galaxies and BGG position. In this case, we investigate the X-ray emission from the system using different scales and visually inspect the X-ray contours alignment around  the position of the BGG, then define the X-ray centre for the group.  In many cases, two groups overlap along the line of sight, the combined data of the Chandra and XMM-Newton allow us to easily distinguish the distinct X-ray centres. In the lower left panel of Fig. \ref{offset_xray}, we show two groups at $z=0.342$ (group 30311) and $z= 0.248$ (group  30224) where  X-ray emissions from these systems overlapping in the line-of-sight. However, the X-ray resolution allows us to define the X-ray centre independently. Depending on the separation of two sources, we assign Flag 1 or 2 to these sources.\\
Flag 3 represents a group which has its own specific X-ray centre but with no spectroscopic members and its redshift is defined based on the photometric redshift of galaxies. Flag 4 corresponds to an extended X-ray source with multiple optical counterparts and it is not possible to determine the contribution of each optical counterpart to the observed X-ray emission. In this case, we define the redshift by considering the position of the bright group galaxies and number of the spectroscopic members. For further detail, we refer the reader to \cite{George11,Gozaliasl14}. 

Column 6 lists group's  $M_{200}$ with $\pm1\sigma$ error in the $[\times 10^{12}\; M\sun]$ units. $M_{200}$ corresponds to the total mass of groups within $R_{200}$ with respect to the critical density of the universe. $M_{200}$ is measured using the $L_X-M_{200}$ scaling relation of \citet{leauthaud2009weak}. Column 7 presents the 0.1-2.5 Kev rest frame X-ray luminosity ($L_X$) with the error in [$\times10^{42}\;erg\; s^{-1}$] within $R_{500}$. Column 8 reports group $R_{200}$ in degrees which is estimated using Eq. 1.\\
In column 9, we report the IGM temperature with corresponding $\pm1\sigma$ error in keV units obtained using $L_{x}-T$ scaling relation. Column 10 presents the cluster/group X-ray flux in the $0.5-2\; keV$ band within $R_{500}$ in units of $[\times10^{-15}\;erg cm^{-2}\; s^{-1}]$ with the corresponding $\pm1\sigma$ errors. Column 11 provides the significance of the X-ray flux estimate which is defined as the ratio of the X-ray flux to its error. Column 12 presents the type of the group redshift: 1) `spec': we  determine groups' $spec-z$  using at least three $spec-z$  members within $R_{500}$. 2) `spec*':  the number of the $spec-z$ members of these groups within $R_{500}$ is less than three members, we thus estimate their $spec-z$ including $spec-z$ members within $R_{200}$. 3) `phot', the redshift of these groups are determined using the photometric redshift of group galaxies.
\begin{figure}
\includegraphics[width=0.495\textwidth]{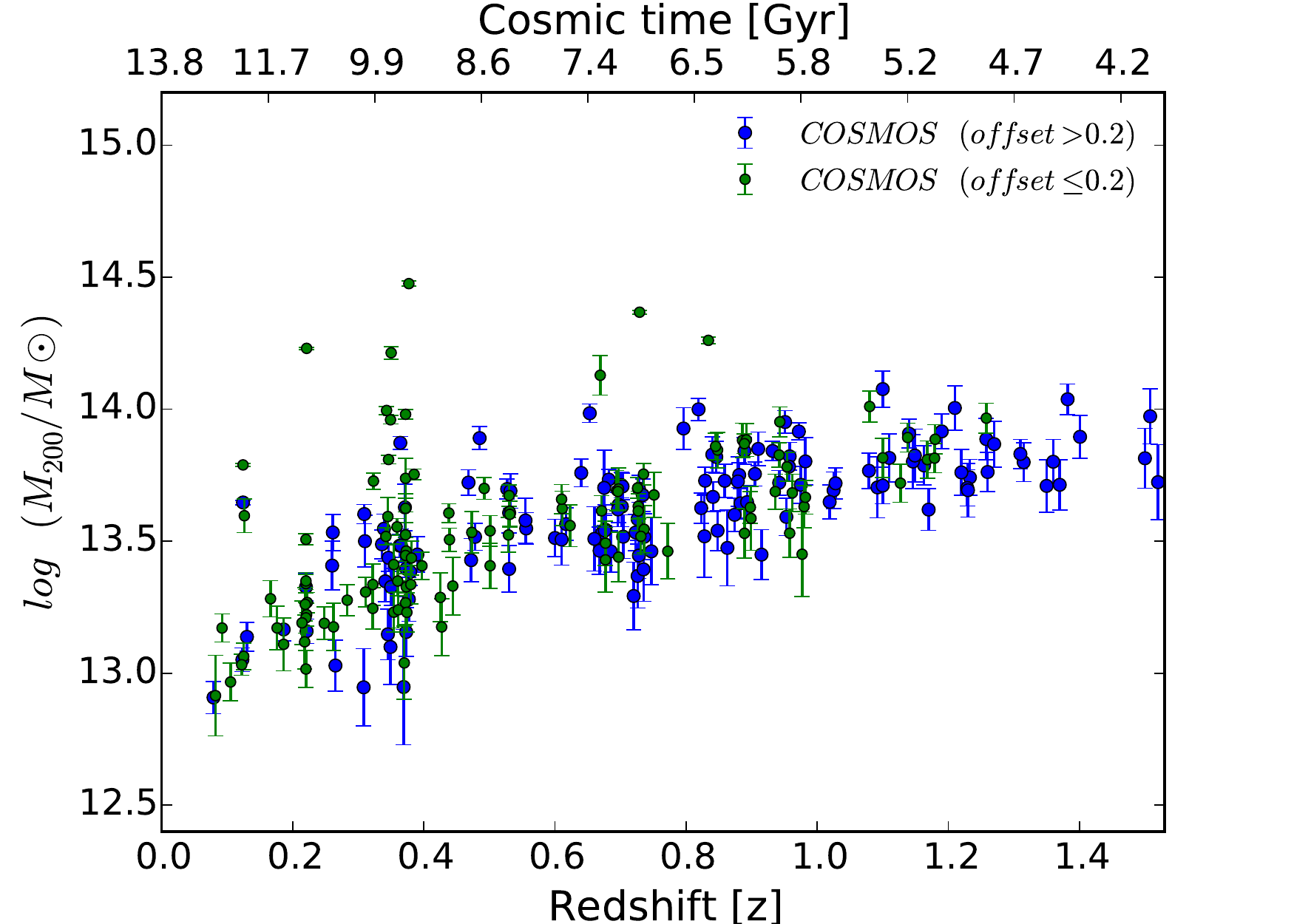}
 \caption{The halo mass of X-ray galaxy groups ($M_{200}$) within $R_{200}$ (defined as the radius delimiting a sphere with interior mean density 200 times the critical density) as a function of redshift. Each colour of data points represents a sample of groups which have been selected based on the ratio of the offset of the BGG position from the X-ray centre of haloes to the $R_{200}$ radius.}\label{mhz}
\end{figure}
\begin{figure*}
\includegraphics[width=0.49\textwidth]{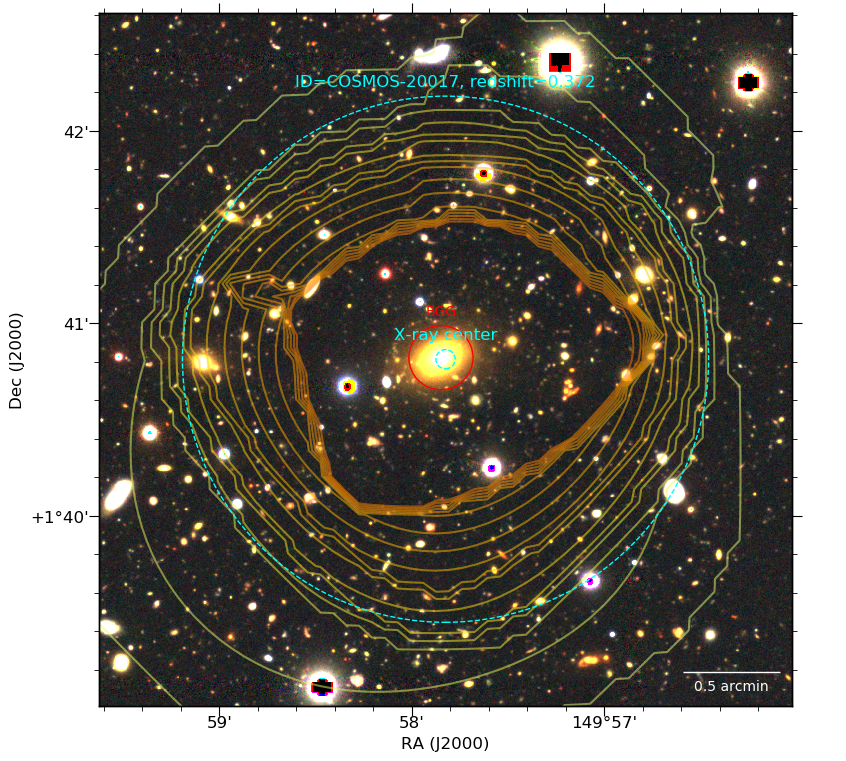}
\includegraphics[width=0.49\textwidth]{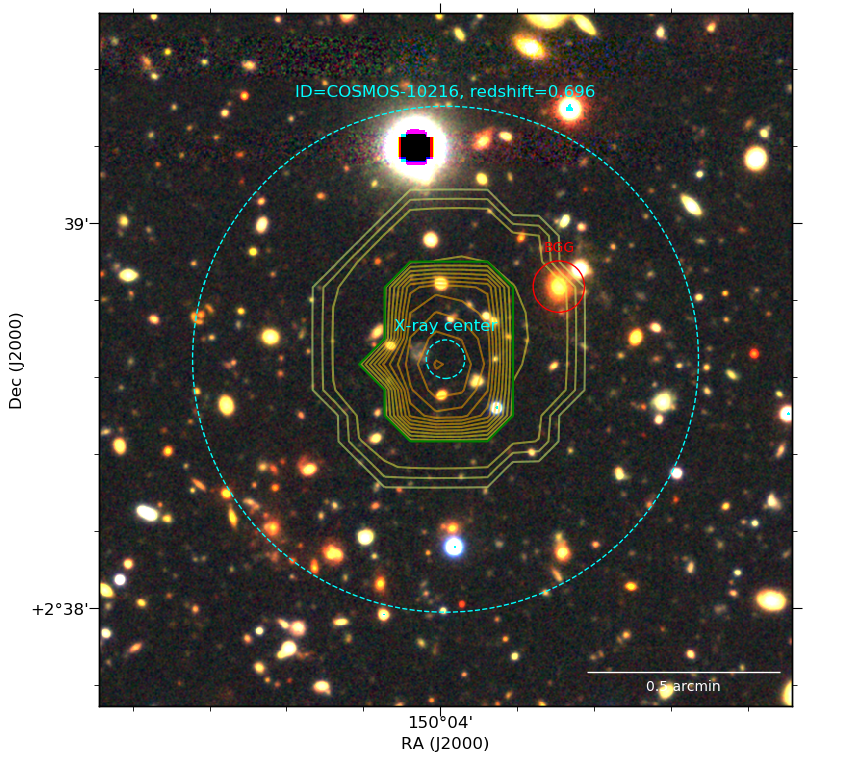}
\includegraphics[width=0.49\textwidth]{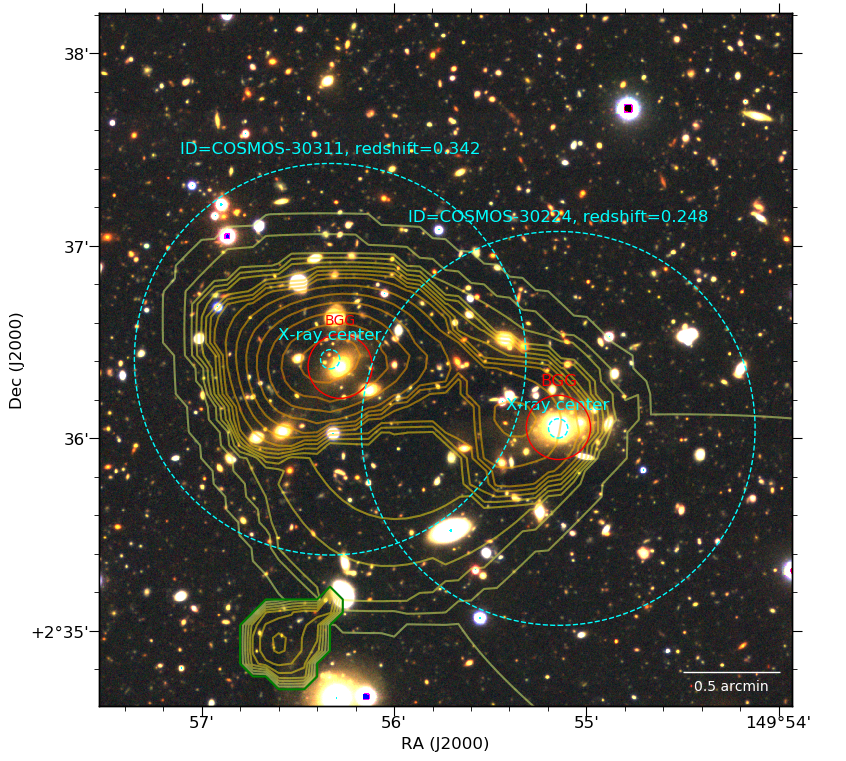}
\includegraphics[width=0.49\textwidth]{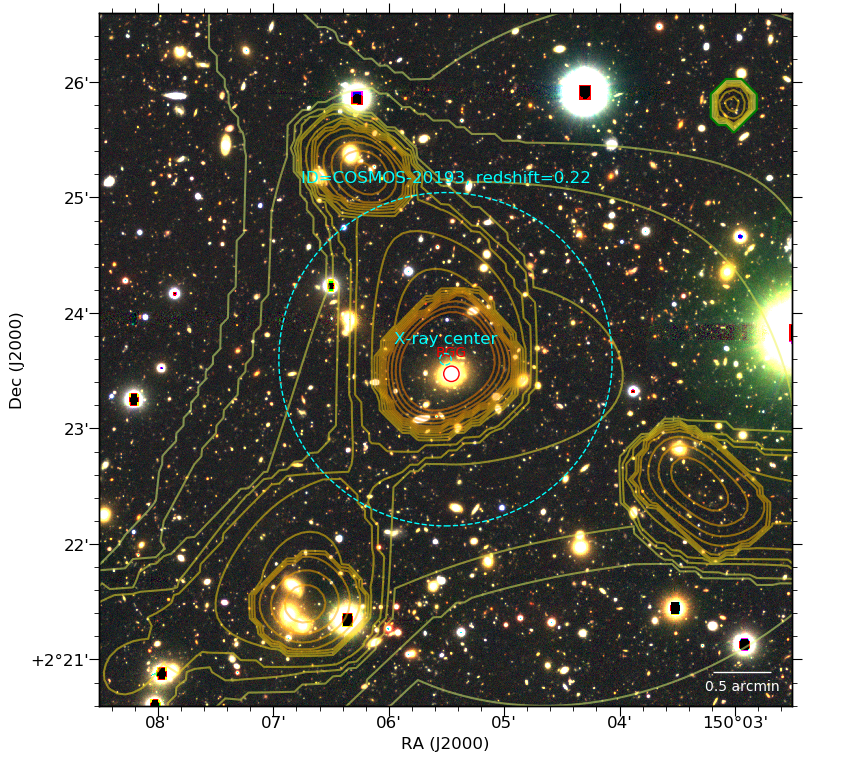}
\caption{The combined RGB optical images of $i, r$, and $g$ Subaru broad bands of four X-ray galaxy groups in COSMOS. The combined Chandra and XMM X-ray emission contours are shown in yellow. Group X-ray centre/peak and $0.5R_{200}$ radius are illustrated with dashed small and large cyan circles. The BGG is marked  with a red circle in each group. \textit{Upper left panel}:  an example of a centrally dominant BGG with no offset from  the X-ray centre of a flag =1 group at $z=0.372$. \textit{Upper right panel}: an example of a BGG with a large offset from the X-ray centre of a flag=1 group at $z=0.696$. \textit{ Lower left panel:} an example of two BGGs with no offset from the X-ray centres of two groups at $z=0.342$ and $z=0.248$ which their extended X-ray emissions are projected. However, the combined X-ray data of Chandra and XMM allows to assign an X-ray centre to each group correctly. \textit{Lower right panel:} an example of BGG with no offset from the group X-ray centre of a group at $z=0.220$. The X-ray contours show three sub-haloes which belong to the parent halo at the centre of the image.}\label{offset_xray}
\end{figure*}

\section{SAMPLE DEFINITION AND DATA}\label{sample}
\subsection{SAMPLE DEFINITION}
 In this section, we  describe the BGG selection and the definition of subsamples. We  make use of our revised catalogue of 247 X-ray galaxy groups with  $M_{h}\sim5\times10^{12} $ to $10^{14.5}M_{\odot} $ at $0.08<z<1.53$, detected  from the  COSMOS field.   
Figure \ref{mhz} presents the halo mass of groups as a function of their redshift. The halo mass ($ M_{h}$) corresponds to the total mass of groups as:
\begin{equation}
M_{\Delta}=\dfrac{4\pi}{3} \Delta \times \rho_{crit}\times R_{\Delta}^3
\end{equation} \label{m200}
where $\rho_{crit}$ is the critical density of the universe and $R_{\Delta}$ is defined as the radius delimiting a sphere whose interior mean density is  $\Delta$ times the critical density of the universe at the group and cluster redshift. Several choices
of $\Delta$ are in use in different studies, from an overdensity
of  180, 200, and 500 times the mean/critical density of  the matter in the Universe \citep{diaferio2001spatial,kravtsov2004dark}.

N-body simulations suggest that  clusters and groups 
are expected to be virialized within over-densities with  $\Delta\sim 200$ times the critical/mean density of the matter in the Universe \citep{Cole:1995ep}. $R_{\Delta=200}$ (hereafter $R_{200}$) is generally used as the characteristic radius to determine cluster/group membership and corresponding physical properties of haloes. In this study, we select group members and BGGs (the most massive and luminous group members) within $R_{200}$ \citep[e.g.,][]{lin2004k}. 

The halo mass of groups in this study corresponds to the total mass of groups ($M_{200}$) within $R_{200}$. We determine $M_{200}$ using the scaling relation of $L_X-M_{200}$ with weak lensing mass calibration as presented by \cite{leauthaud2009weak}. The $L_X-M_{200}$ scaling relation of the COSMOS galaxy groups and similar datasets have been already extensively studied and full details of this relation, and $L_X-\sigma$ and $M_{X}-M_{dyn}$ relations can be found in studies by \cite{leauthaud2009weak,connelly2012exploring,erfanianfar2013x,kettula2015cfhtlens}. Note that we include a $0.08\;dex$ extra error in halo mass estimate, which corresponds to log-normal scatter in the $ L_x-M_{200} $ relation 
\citep{allevato2012occupation}. 
We also study the scaling relations of our sample of $spec-z$ groups in Gozaliasl et al. (in prep.).

Figure \ref{mhz} presents the halo mass of groups ($M_{200}$) within $R_{200}$ (defined as the radius delimiting a sphere with interior mean density 200 times the critical density) as a function of redshift. As can be seen, a large fraction of the groups have a halo mass range of $13.50 < log(\frac{M_{200}}{M_{\odot}}) \le 14.02$.

This mass regime exactly corresponds to a transition zone from  massive clusters to low mass groups which is the main
point of interest in this study. Following \cite{gozaliasl2016brightest,gozaliasl2018brightest}, we select five subsamples of galaxy groups according to the halo mass and redshift plane as:\\ 
(S-I)  $0.04 <$ z $< 0.40$ $ \& $ $12.85 < log(\frac{M_{200}}{M_{\odot}})  \le 13.50 $ \\
(S-II) $0.10 <$ z $\leq 0.40$ $ \& $  $13.50 < log(\frac{M_{200}}{M_{\odot}}) \le 14.02 $\\
(S-III) $0.40 <$ z $\leq 0.70$ $ \& $  $13.50 < log(\frac{M_{200}}{M_{\odot}}) \le 14.02 $\\
(S-IV) $0.70 <$ z $\leq 1.00$ $ \& $  $13.50 < log(\frac{M_{200}}{M_{\odot}}) \le 14.02 $\\
(S-V) $1.00 <$ z $\leq 1.30$ $ \& $  $13.50 < log(\frac{M_{200}}{M_{\odot}}) \le 14.02 $  \\

The subsample of S-II to S-V have the same halo mass range but they are at different redshift ranges. This allows us to compare the stellar mass distribution of galaxies within haloes of the same mass at different redshifts. On the other hand, S-I and S-II have similar redshift ranges but different  halo mass ranges. This enables us to explore the impact of halo mass on the BGG mass distribution over $ z<0.4$. In Tab. \ref{mhmser}, we report the mean statistical and systematic errors in the halo mass of groups  and the stellar mass of BGGs in each subsample.

 \subsection{THE SEMI-ANALYTIC MODELS}
 We interpret our results using two SAMs by \cite{Guo11} (hereafter G11); 
\cite{Henriques15} (hereafter H15). Both models are based on merger trees from the Millennium Simulation \citep{springle2005cosmological} which provides a description of the evolution of the distribution of
dark matter structures in a cosmological volume. While G11 use the simulation in its original
WMAP1 cosmology, H15 scales the merger trees to follow the evolution of large scale structures expressed for the more recent cosmological measurements and Planck results. With respect to the treatment of baryonic physics, G11 and H15 follow the Munich model, L-Galaxies. A summary of the properties of these SAMs are found in \cite{Gozaliasl14,gozaliasl2018brightest} and for their full details, 
we refer readers to \cite{Guo11,henriques2013simulations,Henriques15}.

Both SAMs of H15 and G11 define a parameter known as "type" with different values (0,1,and 2) to select central/satellite galaxies. Type=0 If a galaxy is at the centre of the friend-of-friend (FOF) group, type=1 if the galaxy is at centre of the sub-halo but not at the centre of its FOF group, finally, if a galaxy is a satellite that has lost its sub-halo then its type is 2. We select BGGs in models similar to that in observations and assume them to be the most massive galaxy within the $R_{200}$ of FOF group. 

In this study, we use the data from the H15 and G11 SAMs and randomly select 5000 BGGs within haloes with the halo mass and redshift ranges  as described for S-I to S-V  and  compare the model predictions with our observational results in \S\;5.2 and \S\;5.3.
\subsection{THE HYDRODYNAMICAL SIMULATION OF MAGNETIUM}
For the comparison with hydrodynamical simulations, we use galaxies and galaxy
clusters selected from the Magneticum Pathfinder ({\small www.magneticum.org})
simulation set, which adopts a WMAP7 \citep{Komatsu11} $\Lambda$CDM
cosmology with $\sigma_8 = 0.809$, $h = 0.704$, $\Omega_m = 0.728$,
$\Omega_\Lambda = 0.272$, $\Omega_b = 0.0456$, and an initial slope
for the power spectrum of $n_s = 0.963$.
This suite of fully hydrodynamic cosmological simulations
comprises a broad range of simulated volumes, where for our purpose we choose
the {\it Box2/hr} which uses $2\times1584^3$ particles to simulate a 
cosmological volume of $(500~Mpc)^3$. In this simulation, the stellar component
is represented by stellar sink particles with a mass of
$m_* = 3.5\times10^7M_\odot$ and a gravitational softening of
$\epsilon_*=2.0h^{-1}kpc$. 

All simulations of the Magneticum Pathfinder simulation suite are performed
with an advanced version of the tree-SPH code P-Gadget3 \citep{springle2005cosmological}.
They include metal-dependent radiative cooling, heating from a uniform
time-dependent ultraviolet background, star formation according to \cite{springel2003cosmological}, and the chemo-energetic evolution of the
stellar population as traced by SN Ia, SN II, and AGB stars, including
the associated feedback from these stars \citep{tornatore2007chemical}.
Additionally, they follow the formation and evolution of super-massive
black holes, including their associated quasar and radio-mode feedback.
For a detailed description, e.g. see Dolag et al. (in prep) and \cite{hirschmann2014cosmological,teklu2015connecting}.

We use the {\sc SUBFIND} algorithm \citep{springel2001populating,dolag2009substructures}
to define halo and sub-halo properties. {\sc SUBFIND}
identifies sub-structures as locally over-dense, gravitationally bound
groups of particles which can be associated with galaxies. This
implies that the stellar mass of the main galaxy within a galaxy cluster
or group always represents not only the BCG but also the  intra-cluster light (ICL) component.
The predicted stellar mass function of the simulations generally 
compares well with the observed one over a large range of redshift intervals
(see Gozaliasl et al, in prep). 
Despite this reasonably good agreement, the simulations predicted stellar
masses of BCGs are significantly larger than the observed ones. 
This can be partially a sign of there being still not efficient enough AGN feedback in the 
simulations, but also can be caused by the fact that the stellar masses 
estimates from the BCGs in the simulations account also for the ICL. 
Distinguishing between the BCG and ICL is a notoriously difficult task. 
Based on a dynamical separation of this two stellar components, simulations 
indicate that the stellar mass associated with the BCG is only $\approx45\%$ 
of the total, BCG + ICL mass, see  \cite{dolag2010dynamical,remus2017outer}.
However, it is observationally significant that some fraction of the ICL will
contribute to the observed light from the BCG and therefore the observed
fraction will be larger and depend on the magnitude cut used (see \cite{cui2013characterizing}).
Therefore we assume that the observed stellar mass fraction of the simulated
BCG would generally correspond to $70\%$ of the mass of the total
stellar mass (BCG + ICL) inferred from the simulations.

\subsection{THE BGG SELECTION}
In the current study, the BGG is defined as:
(i) the most massive galaxy,
(ii) within $R_{200}$ of the group X-ray centroid,
(iii) with a redshift that agrees with that of the hosting group
as estimated from all the redshifts available around the X-ray
centroid.

For selecting BGGs, a different choice of apertures from the  the group X-ray centroids are examined.
As mentioned above,  we also examine  different choices of the BGG  selection  within a variety of group radii and apertures from group X-ray centroids ($R_{200}, R_{500}$, and $0.5R_{500}$). We find that when selecting BGGs within $0.5R_{500}$, a number of low mass galaxies are selected as BGGs while there are more massive galaxies at about  $\sim100-300$ kpc from these galaxies.  The differences between the BGG selections within  $R_{200}$ and $R_{500}$ are not meaningful. For the present study, we select BGGs within $R_{200}$, while we show the stellar mass distribution  for all BGGs selected within three different radii in \S\;4.1.\\
Altogether 188 BGGs are selected using their spectroscopic redshifts and 59 BGGs in our  sample are selected using the photometric redshifts considering a $\pm0.007(1+z)$ photo-z accuracy \citep{mccracken2012ultravista,Ilbert13,laigle2016cosmos2015}. All BGGs are visually inspected using the RGB image of hosting groups including  the overlaid extended X-ray emission contours (see Fig. \ref{offset_xray}).
 \begin{table}
 \caption{The average systematic error (se) and the statistical error on the mean (sem) of $ log(M_\star/M_\odot) $ and $ log(M_{200}/M_\odot) $ for S-I to S-V. The error values are in dex.}
 \begin{tabular}{lllll}
 \hline
 \hline
 	sample & $ M_\star$(se) & $ M_\star$(sem)  & $ M_{200} $(se) & $ M_{200} $(sem)\\
 	S-I & 0.12 &  0.08 & 0.16 & 0.01 \\
 	S-II & 0.15 &  0.07 & 0.12 & 0.02 \\
 	S-III & 0.16 &  0.07 & 0.15 & 0.01 \\
 	S-IV & 0.15 &  0.05 & 0.15 & 0.01 \\
 	S-V & 0.19 &  0.05 & 0.16 & 0.02 \\
\hline\hline
 \end{tabular} \label{mhmser}	
 \end{table}
 
\subsection{THE BGG OFFSET FROM HALO X-RAY CENTRE}\label{Sec:BGG-offset}
\subsubsection{Definition of the BGG offset}\label{Ssec:offset def.}

The majority of the brightest cluster galaxies (BCGs) mostly  lie at  the bottom of the potential well of
the host cluster and their X-ray peaks/centres with no considerable offsets\citep[e.g.,][]{jones1984structure,postman1995brightest,lin2004k,lavoie2016xxl}.  It is still unknown whether environments act on galaxy evolution when transiting from a massive cluster-regime to a low mass group-regime. Are BGGs located at the centre of hosting haloes and do they evolve in the same way as brightest cluster galaxies? The multiband data and our well controlled  statistical sample of X-ray groups enable us to answer this question and explore the evolution of BGGs in the galaxy cluster-group transition regime ($M_h\sim10^{13-14}\;M_\odot$) out to the elusive high redshift of $z=1.5$.

It is well-recognized that the dynamical state of groups differs significantly from that of clusters. Groups are likely located where the velocity dispersion of galaxies are sufficiently low which allows merging and interaction between galaxies to happen frequently, thus groups can evolve significantly in a fraction of Hubble time.

N-body modellings experiments have shown that a fraction of current-epoch groups, even those with apparently short dynamical times ($\sim 0.1\; H^{-1}_0$), are probably relatively young systems which might only just been collapsed and they will possibly undergo significant dynamical evolution \citet{barnes1985dynamical}. 

On the other hand, according to the $\Lambda CDM$ model,  groups are the building blocks of massive structures in the universe and they are accreted by massive clusters. Those groups undergo major mergers  and halo mergers, BGGs might be far from the minimum of the potential wells and they could lie far from the X-ray centres (peaks). If this scenario is true, we expect that a fraction of BGGs may offset from the group X-ray peaks even at distances larger than $0.5\;R_{500}$. The brightest cluster galaxies are mostly selected within a small distance to the cluster X-ray centres (see \citep{lavoie2016xxl,harvey2017detection,trevisan2017finer,lopes2018optical,golden2018impact}). In this study, we investigated the BGG offset with respect to three different group radii: $0.5R_{500}$, $R_{500}$, and $R_{200}$. Therefore, the offset of BGGs is defined as the ratio of their angular separations from the halo X-ray centroid (r) to the given group's radius (e.g., here $R_{200}$) as:\\ 
\begin{equation}
offset=\dfrac{r}{R_{200}}.
\end{equation}

We select a clean sample of the observed groups (flag$\leq 3$) and BGGs in COSMOS where we are able to assign an X-ray centre for each group and study the BGG offset with excluding groups with low identification- and X-ray flux-significance. 

Based on the  BGG offset from the X-ray centroids, we classify them into the following three classes in each subsample (S-I to S-V): \\
(i) Central dominant BGGs with offset$\leq 0.2$, \\
(ii) Large offset BGGs with offset$>0.2$, and\\
(iii) All BGGs with any offset$\leq 1$.\\
In Fig. \ref{mhz}, we   illustrate  haloes having BGGs with low and large offsets with filled green and blue circles, respectively.
In the upper panel of Fig. \ref{offset_xray}, we also show two examples of the central dominant BGG (left-hand) and BGG with large offset (right-hand). 

In the following subsection we investigate the distribution and evolution of the BGG offset and its relations with the halo mass, X-ray flux, the significance of the X-ray flux, and r-band magnitude gap between two brightest group galaxies.

\subsubsection{Distribution of the BGG offset}\label{Ssec:offset-dist}
\begin{figure*}
\includegraphics[width=0.49\textwidth]{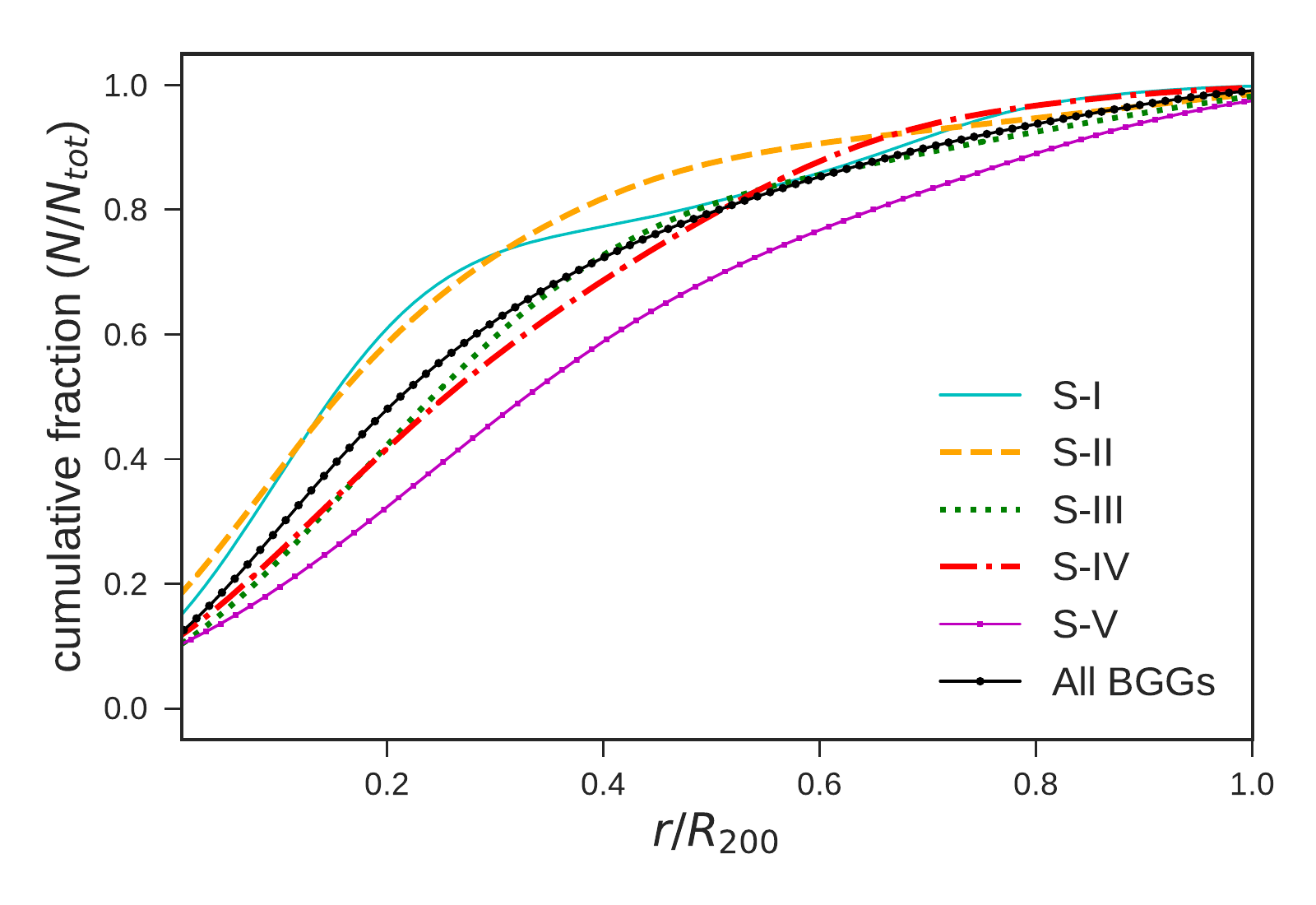}
\includegraphics[width=0.49\textwidth]{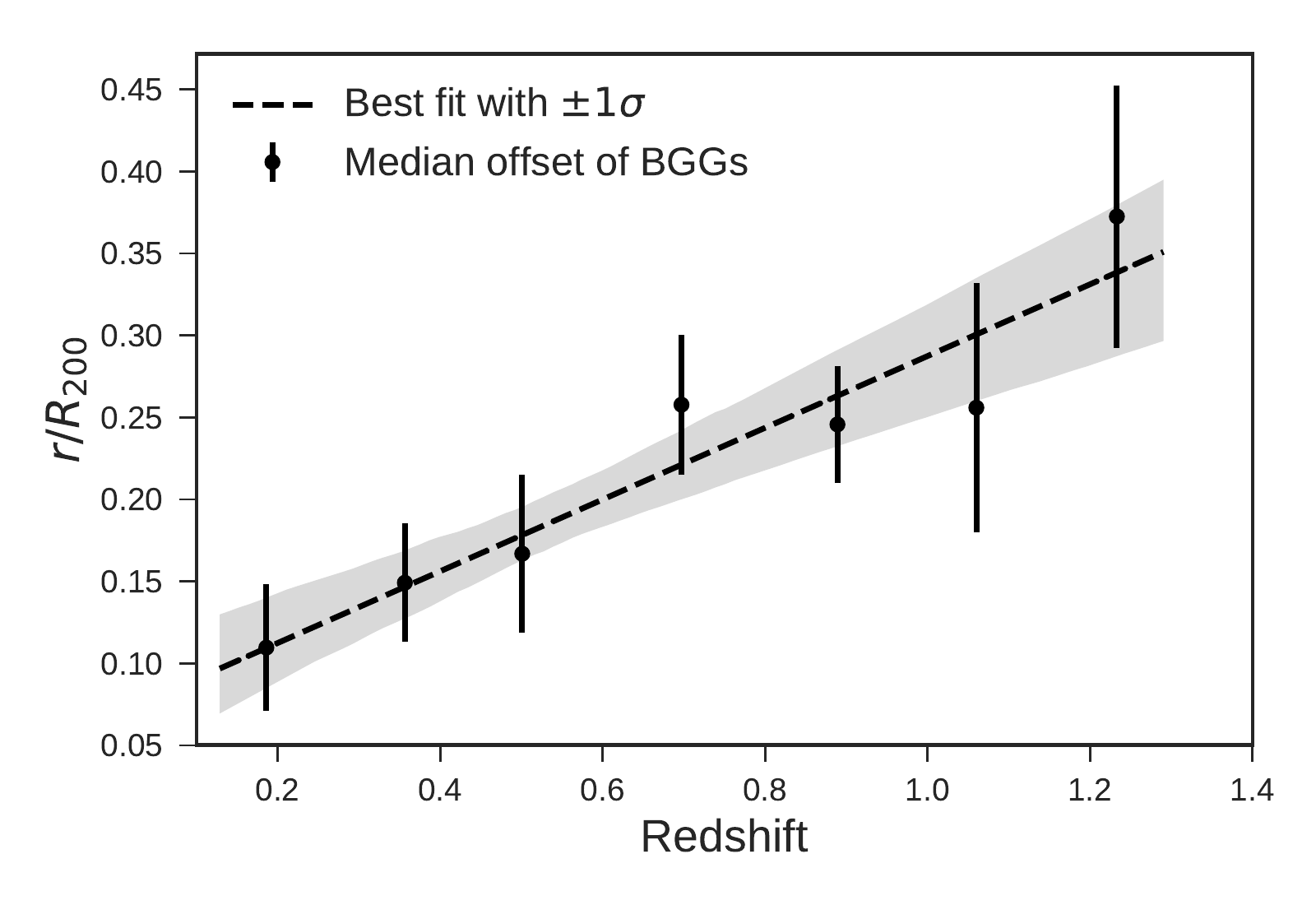}
\includegraphics[width=0.49\textwidth]{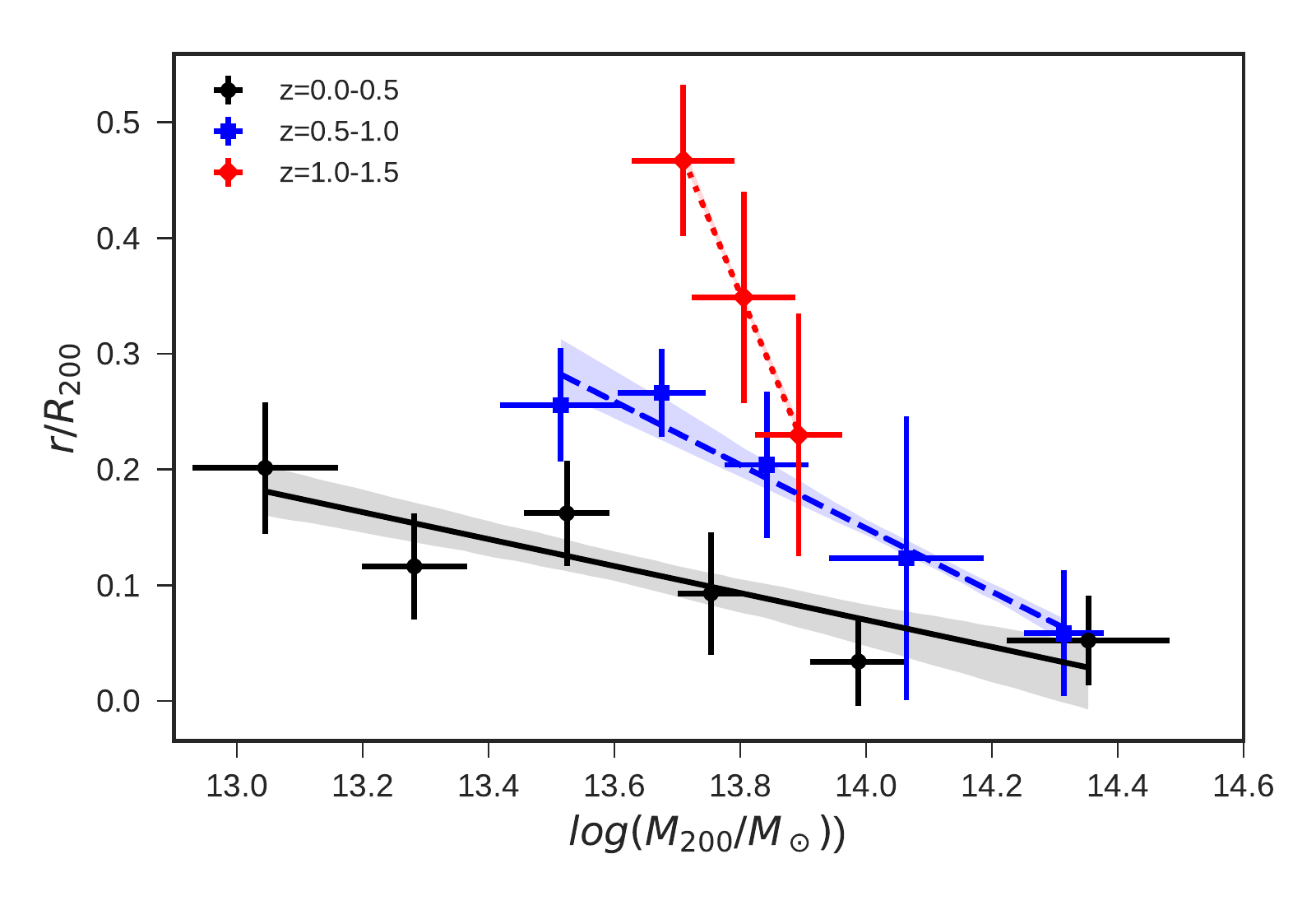}
\includegraphics[width=0.49\textwidth]{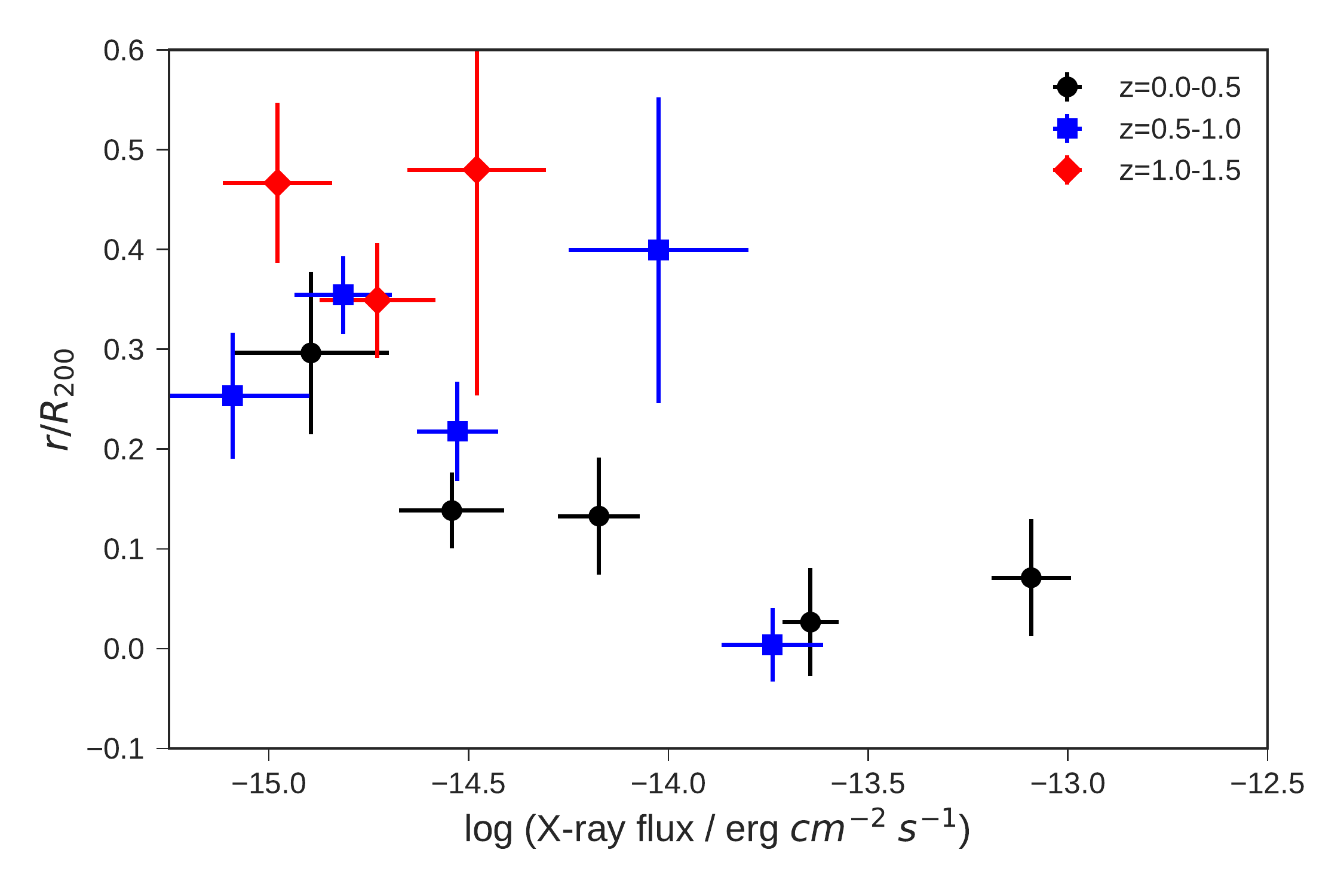}
\includegraphics[width=0.8\textwidth]{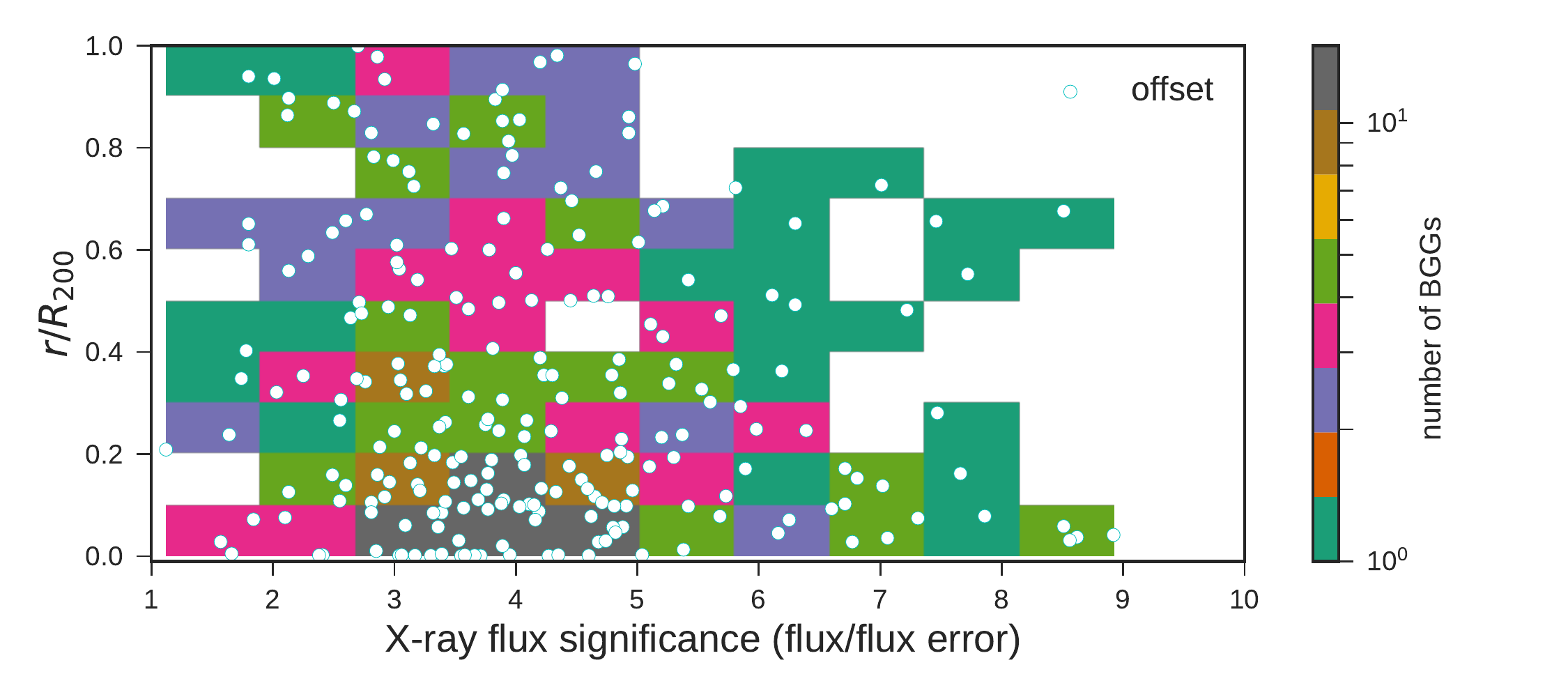}

 \caption{\textit{Upper left panel:} Cumulative distribution of the projected radial  offset of the BGGs from the X-ray centre for S-I to S-V and the full sample of BGGs (solid black line with filled circles). The distance of the BGG from X-ray centre (r)
is normalized to the radius of each group, $R_{200}$. The number of galaxies located within a given distance is normalized to the total number of BGG within each subsample. \textit{Upper right panel:} The redshift evolution of the median values of the offset of the full sample of BGGs from the X-ray centre of hosting groups. The BGG offset from X-ray centre decreases as a function of the redshift  by a factor of 3 since $z=1.3$ to present day. \textit{Middle left and right panel:} The BGG offset ($r/R_{200}$) as a function of the group's halo mass ($M_{200c}$) (left) and the group X-ray flux (right) for BGGs at $z=0.0-0.5$ (solid black line and filled circles), $z=0.5-1.0$ (dashed blue lines and filled squares), and $z=1.0-1.5$ (dotted red line and filled diamonds), respectively. The BGG offset decreases as a function of increasing halo mass and the slope of this relation increases interestingly  with increasing redshift. \textit{Lower panel:}  The BGG offset ($r/R_{200}$) as a function of the flux significance (flux/fluxer). The colour bar presents the number of BGGs  within given offset and the flux significance bins. }\label{off}
\end{figure*}

The upper left panel of Fig. \ref{off} shows the cumulative distribution of the projected distance from the X-ray peak for the full sample of BGGs (black dot-line) and BGGs for S-I to S-V. We see that
$\sim80\%$ of all BGGs (black line) have an offset $\leq 0.5 R_{200}$. We are aware that the adoption of a large radius ($R_{200}$) could result in a larger background/foreground contamination in selecting BGGs. This
is a major problem if there is no complete spectroscopic coverage. We also consider that for merging systems
$R_{200}$ could be overestimated, implying a BCG selection
within an unnecessarily large radius. To avoid these uncertainties, we first exclude groups with low identification significance applying flag$\leq 3$ limit, visual inspection of the group's image, and  then examine the BGG selection within smaller radii ($0.5R_{500},R_{500}$). In addition, our groups are mostly spec-z systems with spec-z BGGs. Finally, we find that it is true that adopting smaller radii may decrease the background/foreground contamination, however, it causes that in some groups very low-mass centrally located  satellite galaxies with $M_*<10^{9}M_\odot$ are selected as BGGs, while the true BGGs lie with offset from the group X-ray centroid. We investigate this in more details in $\sec$ \ref{sm-offset} 

 We suggest that for a sample of photo-z groups with not enough spec-z using a smaller radius like $0.5 R_{200}$ to select BGGs could be a confident approach, so that you can choose the true BGGs, not risking background/foreground contamination at larger radius.

We apply the K-S test and quantify p-value and the differences between the  cumulative distributions of the BGG offset for S-II ($z=0.1-0.4$) and S-V ($z=1.0-1.3$). The p-value corresponds to $\sim0.06$ and we are not able
to reject the null hypothesis that the two samples were drawn from the
same distribution. There is also a $D=0.36$ difference between two distributions at $r/R_{200}=0.2$. As a result, we suggest that BGGs are likely to become central galaxies with decreasing redshifts and  their offsets  widen at higher redshifts.
\subsubsection{Evolution of the BGG offset}\label{Ssec:offset-z}
The upper right panel of Fig. \ref{off} presents the median value of the BGG offset from the X-ray centre as a function of the redshift of the host groups. We note that the trend of the $r/R_{200}-z$ relation does not change significantly when we exclude the BGGs for S-I from the data and plot this relation for S-II to S-V. This redshift evolution of the BGG offset is  evident here and we quantify the relation between the BGG offset ($r/R_{200}$) and redshift (z) as: 
\begin{equation}
r/R_{200}=(0.174\pm0.002)+(0.167\pm0.003)\times z
\end{equation}
we find that the BGG offset ($r/R_{200}$) evolves as a function of redshift and it decreases by $\sim0.25$ from $z=1.53$ to the present day.

\subsubsection{The relation between the BGG offset and halo mass}\label{Ssec:offset-mh}
We show the median value of the BGG offset as a function of the group's $M_{200c}$  in the middle left  panel  of Fig. \ref{off}.  The BGG offsets are plotted as a function of the group's critical halo mass $log(M_{200c}/M_\odot)$  for the given three redshift bins, $z=0.0-0.5$ (solid black line and filled circles), $z=0.5-1.0$ (dashed blue line and filled squares), and $z=1.0-1.5$ (dotted red line and filled diamonds), individually. We find that the BGG offset decreases with increasing halo mass as the slope of the relation negatively increases as a function of increasing redshift. The slope of this relation for each redshift bin, $z=0.0-0.5$, $z=0.5-1.0$, and $z=1.0-1.5$ is quantified using the linear regression model as $-0.116\pm0.002$, $-0.273\pm0.04$, and  $-0.922\pm0.188$, respectively.  
\subsubsection{The relation between the BGG offset and the group X-ray flux}\label{Ssec:offset-flux}
As discussed above, the BGG offsets widen towards higher-z. This could be due to lower fluxes
and SNR ratios or as a result of a group evolution. Hence, to address how much of the effect could be driven by noise, we inspect the relationship between the median value of the BGG offset and the group's X-ray flux (see the middle right panel  of Fig. \ref{off}) at three redshift bins, $z=0.0-0.5$ (solid black line and filled circles), $z=0.5-1.0$ (dashed blue line and filled squares), and $z=1.0-1.5$ (dotted red line and filled diamonds), respectively. It appears that the data shows a weak negative correlation between the BGG offset and the group X-ray flux at $z=0.0-0.5$ while there is no correlation at high redshifts. This finding is also more evident in the lowest panel of Fig. \ref{off} where we present the relation between the offset of the majority of our BGGs sample  and  their host groups' flux significances ($<10$). We define the flux significance as the ratio of the flux to flux error. In this panel, we also show a 2-dimensional histogram which counts BGGs within given offset and flux significance bins.  We find that there is no preferential trend between offset and flux significance and the BGG offset spreads a wide dynamic ranges at any given flux significance. This means that the offset of BGGs from the X-ray centroids are not driven by observational noise.

We note that there are 15 additional systems with higher flux significance (10-62) outside the plot range. These systems are the most massive groups and clusters in our sample and are  hence have very bright X-ray significances. Their BGG/BCG offsets range from $(\sim0-0.4)$.

\subsubsection{The relation between the BGG offset and magnitude gap}\label{Ssec:offset-gap}

The difference between the first and the second ranked galaxy magnitudes in groups is often considered a tracer of their merger histories and dynamical evolution \citep{barnes1989evolution,ponman1994possible,gozaliasl2014gap,raouf2014ultimate,gozaliasl2014evolution,khosroshahi2014optically,trevisan2017finer}. Using N-body simulations
of isolated groups, various studies found early on that
galaxy mergers in groups will lead to runaway growth of the most massive central
galaxy \citep{carnevali1981merging,cavaliere1986dynamical,barnes1989evolution,mamon1992cluster}.

This growth  may occur independently
of the merger mechanism \citep{mamon1987dynamics}, whether the
group evolves through direct merging between galaxies or due to orbital
decay via dynamical friction that causes group galaxies to lose
energy and angular momentum against a diffuse background. In both hypotheses, the growth of the BGG happens at the expense of the second-brightest group galaxy (SBGG), because the merger cross-section  for SBGG  is greater than
that of the less massive and luminous satellites, in addition, the
dynamical friction time scales as the inverse of the galaxy
subhalo mass, leading to faster orbital
decay of SBGG \citep{chandrasekhar1943dynamical}, hence more rapid merging with
the BGG. According to this scenario, as a group evolves through merging, the magnitude gap should also grow in time. The final product of such a rapid growth of the central group galaxy,  is a group that includes a giant elliptical galaxy surrounded with some faint satellites with a luminous X-ray halo (bolometric $L_X > 10^{42}\; h^{-2}_{50}\;erg\;s^{-1}$), exhibiting a large magnitude gap ($\Delta M_{1,2}>2$) with the second SBGG within $0.5R_{200}$ \citep{jones2003nature}. These types of groups are known `fossil groups' and the first fossil groups was discovered by \citep{ponman1994possible}. 

Fossils are early formed and relaxed systems \citep{gozaliasl2014evolution}, as a result,   the BGGs in fossils  are  central dominant galaxies with the lowest offset from the group X-ray-centroid halo. We investigate here the relation between the BGG offset from the x-ray centre and the magnitude gap between two brightest group galaxies.

Recently, \cite{lopes2018optical} used two samples of the Sunyaev-Zel'dovich (SZ) effect- and the X-ray-selected sample of  ($z<0.11$) clusters and estimated the dynamical state of clusters using the BCG-X-ray centroid offset, and the magnitude gap between the first and second BCGs. They recommend an offset cut-off $\sim0.01\times R_{500}$ to separate relaxed and disturbed clusters. Regarding the magnitude gap, the separation can be done at $\Delta m_{1,2} =1.0$. They showed that 20\% and 60\% of the relaxed and disrupted clusters include BCGs with a large offsets.

\cite{trevisan2017finer} studied the magnitude gap and the conditional luminosity function of the SDSS groups \citep{yang2007galaxy} and found that
some groups preferentially small-gap groups have more than one central galaxy. 

\cite{golden2018impact} 
claim the intrinsic scatter in the BCG stellar mass at fixed halo mass can
be reduced if accounting for the magnitude gap. Finally, \cite{harvey2017detection} predict a residual BCG wobbling in clusters due to
previous major mergers.

\begin{figure}
\includegraphics[width=0.49\textwidth]{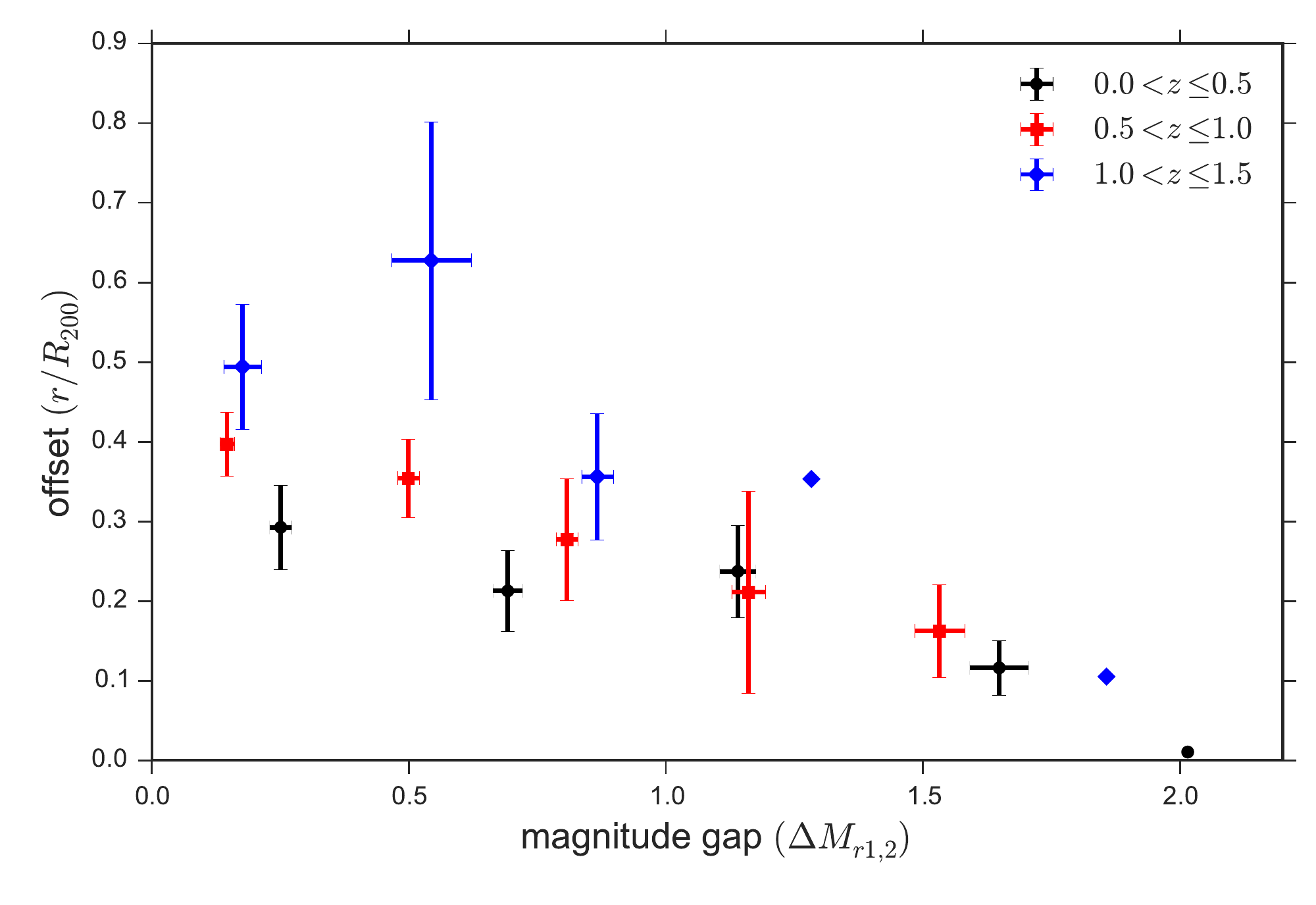}
\caption{The relation between the BGG offset from the group X-ray peak/centre and the r-band magnitude gap ($\Delta M_{1,2}$) between the first and the second brightest group galaxies within $R_{200}$  at $z=0.0-0.5$ (black circles), $z=0.5-1.0$(red squares), and $z=1.0-1.5$ (blue diamonds)). The BGG offset from the X-ray centre of groups shows a negative  correlation with the magnitude gap with no redshift dependent trend. Groups with a large magnitude gaps (e.g., fossils) represent BGGs with the lowest offset from the halo X-ray centre.}\label{FIG:gap_offset}
\end{figure}

Figure \ref{FIG:gap_offset} shows the relation between the BCG offset from the group X-ray peak/centre ($r/R_{200}$) as a function of the magnitude gap between the first and second brightest group galaxies within $R_{200}$. We note that the offset/magnitude gap estimation within $0.5R_{200}$ represent similar trends. We plot the offset-gap relation for groups at three different redshift ranges, $z=0.0-0.5$ (black circles), $z=0.5-1.0$(red squares), and $z=1.0-1.5$ (blue diamonds)). We find that the BGG offset from the halo X-ray centre decreases as a function of increasing magnitude gap with no redshift dependent trend.

In summary, we conclude that the BGG offset depends on the halo mass with a redshift dependent trend. The offset also negatively correlates with magnitude gap of groups, suggesting that the BGG offset is as an important and simple observable which can be used to determine the group dynamical states. This parameter is not driven e.g. by SNR in the observational data, and 
not driven by observational noise. The off-central BGGs probably  reside in groups which are more likely to
have experienced a recent halo merger or are undergoing a merger. The host groups of off-central BGGs generally  include two massive luminous galaxies and they will possibly merge into one and probably get closer to the group's centre, expecting that the BGG offset will decrease with time as the group evolves dynamically.  
\section{DIFFERENCES IN THE STELLAR MASS DISTRIBUTION OF THE CENTRAL DOMINANT BGGS AND THE LARGE OFFSET BGGS}
This paper uses the physical properties of the galaxies from the COSMOS2015 catalogue presented by \cite{laigle2016cosmos2015}. The main improvement in this catalogue compared to previous COSMOS catalogue releases is the addition of new, deeper NIR and IR data from the UltraVISTA and SPLASH projects. The COSMOS2015 catalogue contains precise stellar masses for over half a million objects at the $\sim2\;deg^2 $ 
COSMOS field. Including new ${{YJHK}}_{{\rm{s}}}$ observations from the UltraVISTA-DR2 survey, Y-band observations from Subaru/Hyper-Suprime-Cam, and infrared data from the Spitzer Large Area Survey with the Hyper-Suprime-Cam Spitzer legacy program, this highly optimized near-infrared-selected catalogue allow study of the evolution of galaxies and the environment effects in the early universe. For more details on the stellar mass estimation and the physical properties of galaxies, the reader is referred to \cite{laigle2016cosmos2015}.

\subsection {THE BGG POSITION WITHIN ITS HOST HALO}\label{sm-offset}
 The position of BGG in a group  does not
always correspond to the centre of the group potential well \citep{beers1983environment,zabludoff1993kinematics,lazzati1998location,lin2004k,von2007special,skibba2010brightest,oliva2014galaxy}. This is probably due to recently accreted relatively massive satellites that have not still merged and may still be growing \citep{skibba2010brightest}. These galaxy groups may also recently be dynamically relaxed. For a sample of massive and low-z clusters whose BCGs are generally central galaxies, such a selection may  not affect the scientific conclusion. However, this selection criterion can significantly affect the scientific results reached on BGG properties and evolution. In this study, we use a clean sample of BGGs selected from (flag$\leq3$) groups and examine the BGG selection within three different radii ($0.5R_{500}, R_{500}$, and $R_{200}$). We investigate possible contamination of the BGG sample by the second brightest group galaxy (hereafter SBGG) selected within the mentioned aperture sizes. 

\begin{figure*}
\includegraphics[width=0.88\textwidth]{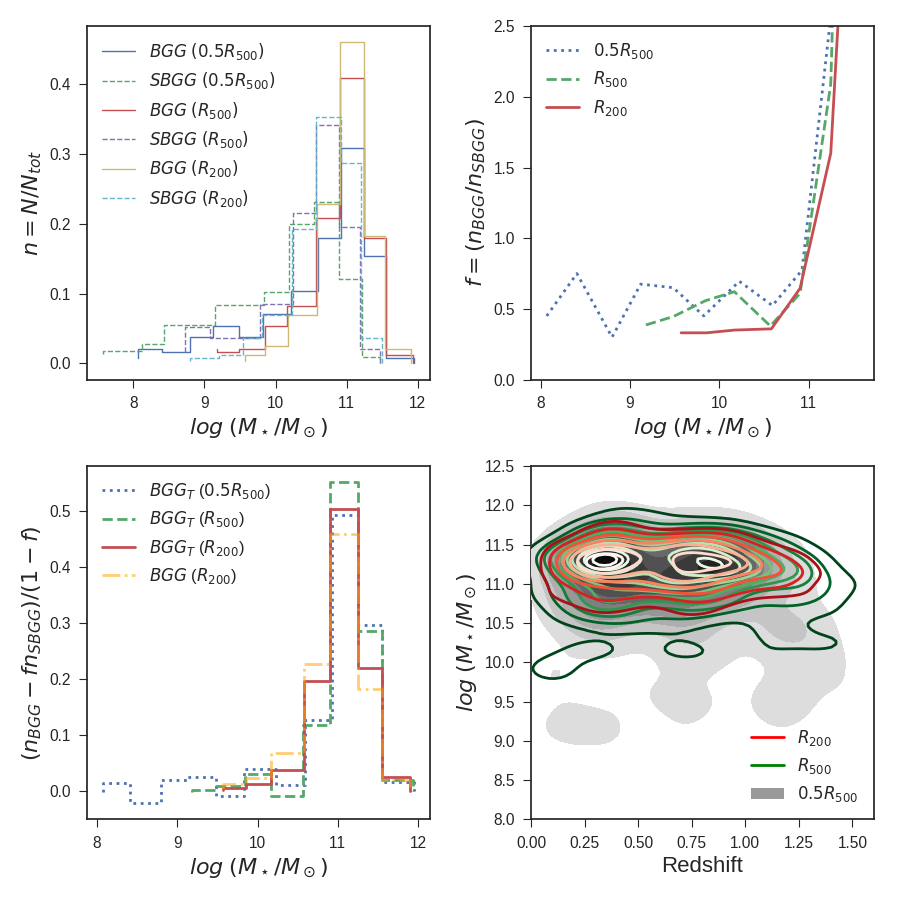}
\caption{\textit{Upper left panel:} The  distribution of $log(M_\star/M_\odot)$ for the full sample of BGGs (solid curves) and the second brightest group galaxy (dashed curves) selected within three different apertures from the group X-ray centre: $R_{200}$, $R_{500}$, and $0.5R_{500}$. It appears the stellar mass distribution of the BGGs tend to skew to lower masses with decreasing the group radius from $R_{200}$ to $0.5R_{500}$. \textit{Upper right panel:} The BGG to SBGG ratio per stellar mass bin. The probability of missing a BGG increases from 40\% to 50\% when the aperture size for BGG selection increasing from $0.5R_{500}$ to $R_{200}$.  \textit{Lower left panel:} The stellar mass distribution of  bona fide (true) BGGs ($BGG_T$) selected within $R_{200}$ (solid red curve), $R_{500}$ (dashed green curve), and $0.5R_{500}$ (dotted blue curve) after subtracting possible contamination by SBGG. The yellow dot-dashed line shows the stellar mass distribution of BGGs selected within $R_{200}$ without   subtraction of any contamination. \textit{Lower right panel:} The density map of the stellar mass versus redshift for BGGs selected within  $R_{200}$ (solid red contours), $R_{500}$ (solid green contours), and $0.5R_{500}$ (shaded grey area). Both panels show that by decreasing the group radius from $R_{200}$ to smaller radii e.g.,  $0.5R_{500}$, where BGGs are selected, a number of low mass galaxies which are located in the central region of groups are chosen as BGGs while there are more massive galaxies a little farther from the group X-ray centre with no dependence on redshift.}\label{sm_r}
\end{figure*}

The upper panel of Fig. \ref{sm_r} presents the stellar mass distribution of BGGs (solid curves) and SBGGs (dashed curves) selected within $0.5R_{500}$, $R_{500}$, and $R_{200}$, respectively. The y-axis corresponds to the normalised count of BGGs/SBGGs ($n=N/N_{tot}$). We find that  distributions of the stellar mass of BGGs and SBGGs corresponding for each radius peak at $log(M_*/ M_\odot)\sim11.1-11.4$ and $log(M_*/ M_\odot)\sim10.7-11.0$, respectively. All distributions tend to skew to lower masses, however, the skewness increases with decreasing the aperture size from $R_{200}$ to  $0.5R_{500}$. In order to quantify possible contamination of the BGG sample by SBGG we measure the BGG to SBGG ratio ($f=n_{BGG}/n_{SBGG}$) for each stellar mass bin. The ratio remains constant at $\sim0.5$ for $0.5R_{500}$ and $R_{500}$ and 0.4 for $R_{200}$ at $log(M_*/ M_\odot)\leqslant10.5$, then it increases for high masses. We find that the BGG selection could potentially be more contaminated at $log(M_*/ M_\odot)\leqslant10.5$ and the probability of missing a true BGG increases from 40\% to 50\% when the aperture size for choosing BGGs decreasing from $R_{200}$ to  $0.5R_{500}$.

The lower left panel of Fig. \ref{sm_r} shows the stellar mass distribution of bona-fide (true) BGGs (hereafter $BGG_T$). We  have subtracted possible contamination of the BGG sample by SBGG. We also show the stellar mass distribution of BGGs ($R_{200}$) without subtracting the contamination. The y-axis corresponds to $(n_{BGG}-f\times n_{SBGG})/(1-f)$. As is seen the stellar mass distribution of bona fide BGGs selected from different apertures are approximately similar at $log(M_*/ M_\odot)>10.5$. In other word, among $BGG_T$s there is no evidence for a BGG mass to be below $log(M_*/ M_\odot)\sim10$, what are below this mass are possibly misidentified BGGs. It is not expected that a galaxy with a stellar mass of $log(M_*/ M_\odot)\sim8-9$ to be as a central group galaxy at $z<1$, what is seen in the $M_*$ distribution of BGGs selected within $0.5R_{500}$. In addition, we find no significant difference between the mass distribution of $BGG_T$s and that of BGGs without subtraction of the contamination.

The density maps of the stellar mass of those BGGs  versus redshift are also plotted in the lower right panel of Fig. \ref{sm_r}. We find that by decreasing the group radius the stellar mass distribution of BGGs tends to skew to lower stellar masses. As an example, when selecting BGGs within  $0.5R_{500}$ (even $R_{500}$), a number of low mass satellite galaxies  with $M_*\sim 10^{8-9}M_\odot$ are chosen as BGGs in some groups while there are more massive bright galaxies at 1-3 hundred $kpc$ in these systems. After a careful inspection of groups and their associated BGGs, we conclude that the best radius for selecting BGGs within our groups is  $R_{200}$ which approximately corresponds to the physical virial radius of haloes.\\
 In order to test whether the stellar mass distribution of BGGs selected within $0.5R_{500}, R_{500}$, and $R_{200}$  are drawn from the same distribution, we apply two-sided K-S test and measure K-S statistics and p-values. Table \ref{Tab:ks} presents the K-S test results. We find that the stellar mass distribution for BGGs selected within $R_{200}$ is slightly different with that of the BGGs selected within $R_{500}$. This difference increases when selecting BGGs within $0.5R_{500}$. However, we cannot reject the hypothesis that the distributions of the two samples of BGGs selected within $R_{500}$ and $0.5R_{500}$ are the same.

\begin{table}
\caption{The results of two-sided K-S  test to examine whether the stellar mass distribution of BGGs selected within $0.5R_{500}, R_{500}$, and $R_{200}$ are drawn from the same distribution. Column 1 presents two samples that are compared. Column 2 and column 3 lists the K-S statistics and p-values. } \label{Tab:ks}
\begin{tabular}{cccc}

BGG samples&K-S statistics& p-value\\
\hline\\
($R_{200},R_{500}$)& 0.095& 0.210\\
($R_{200},0.5R_{500})$& 0.244& 7.138\\
($R_{500},0.5R_{500})$& 0.158& 0.004\\
\hline

\end{tabular}
\end{table}
 \begin{figure*}
\includegraphics[width=0.88\textwidth]{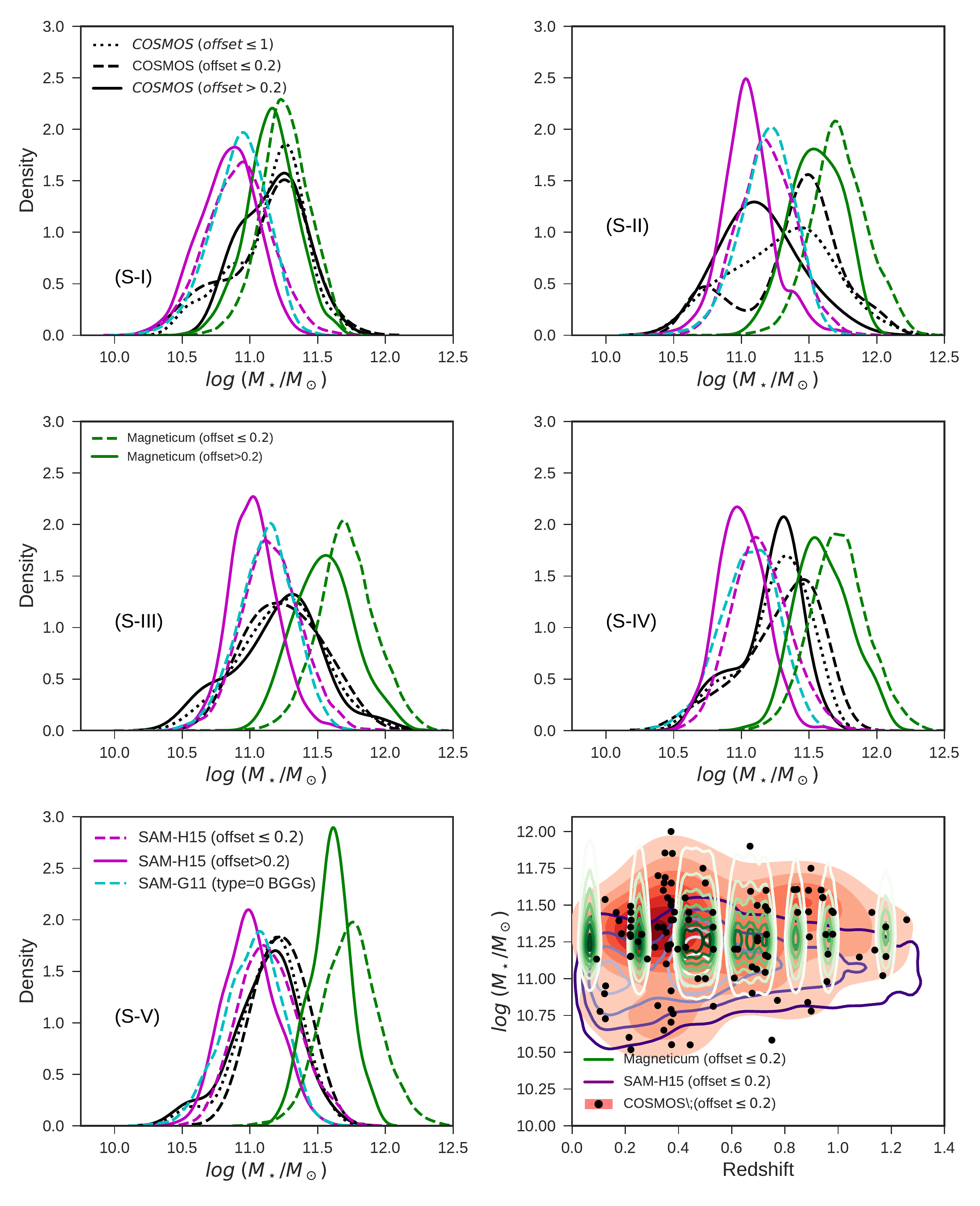}
\caption{The smoothed distribution of $log(M_\star/M_\odot)$ for the full sample of BGGs (dotted black line), BGGs with low offset from the X-ray centre (offset=$\dfrac{r [deg]}{R_{200}[deg]}\leq 0.2$)  (dashed black line), and BGGs with large offset (offset $>0.2$), by Gaussian kernel density estimator. The smoothed stellar mass distribution of the central dominant BGGs and the large  offset BGGs predicted by H15 are plotted with dashed and solid magenta lines. The stellar mass distribution of the central dominant BGGs and the large offset BGGs in the hydrodynamical simulation of Magneticum are plotted with dashed and solid green lines. The dashed cyan  line shows the stellar mass distribution of the central BGGs (type=0 galaxies) in the G11 SAM. The lower right panel presents $log(M_\star/M_\odot)$ as a function of redshift. The density map of BGGs with low offsets in observations are shown with the shaded red area. Black points present the stellar mass versus redshift for the central dominant BGGs (offset $<0.2$) in observations. The purple  and green contours illustrate the density contours for the central BGGs in the H15 model and the Magneticum simulation.} \label{sm}
\end{figure*}
\subsection {DISTRIBUTION OF THE STELLAR MASS OF BGGs}\label{sm}

This study  aims to measure differences between the stellar mass distribution of the central dominant BGGs with offset$\leq 0.2$ with those BGGs with large offset$ >0.2$. We also compare the stellar mass distributions BGGs with low and large offset with the full sample of BGGs without considering offset. 

To decrease the dependence of the stellar mass distribution to bin-size and the end point of histogram, we use the Gaussian Kernel Density Estimation (KDE) technique \citep{rosenblatt1956} and determine the probability density function of $log(M_\star/M_\odot)$. 

Figure \ref{sm} presents the smoothed distribution of $ log(M_\star/M_\odot)$ of BGGs. Each panel, except for the lower-right panel, presents the $log(M_\star/M_\odot)$ distributions for a subsample of S-I to S-V. We show the $log(M_\star/M_\odot)$ distributions for BGGs with low offset, BGGs with large offset, and all BGGs  with a dashed, solid, and dotted black lines, respectively. The distributions of $log(M_\star/M_\odot)$ of the central and the large offset BGGs selected from the H15 SAMs are illustrated with dashed and solid magenta lines. We also show the distribution for the central BGGs from the G11 model with dashed cyan line. The stellar mass distributions for the central and the large offset BGGs in the Magneticum simulation are plotted with dashed and solid green lines.  The main findings of the stellar mass distribution of BGGs for S-V to S-I in Fig. \ref{sm} are summarized as follows:
\begin{enumerate}
	\item For S-V, we find that the stellar mass distribution of the central BGG approximately shows a single Gaussian distribution, while the distribution for BGGs with large offset shows a second peak at around $log(M_\star/M_\odot)=10.5$, and this causes the shape of the stellar mass distribution to deviate from a Gaussian distribution. Overall, we find a good agreement between model predictions and observations and the deviation of the position of the peaks (mean stellar mass) among observation and predictions for central BGGs and those with large offset are not significant. 
    
The Magneticum simulation overestimates the stellar mass of both the central dominant BGGs and the large offset BGGs in observations by $\sim0.6\;dex$. 
    
    \item For S-IV, the shape of $log(M_\star/M_\odot)$ distributions for all BGGs and BGGs with large observational offsets  are similar to those for S-V, however, the height of the second peak seems to increase by a factor of two.  We observe that the distribution for the central BGGs in observations tends to skew to lower masses and the position of the centre of the peak moves to higher masses  compared to the same for S-V distribution and there is a deviance of the stellar mass evolution by $\sim0.2$ dex from $z=1.3$ to $z=0.7$. The mean of the stellar mass of central dominant is also higher than that of  BGGs with a large offset by $\sim0.2$ dex in agreement with prediction by H15. G11 and H15  similarly predict the stellar mass distribution of central BGGs and both model under-predict the mean stellar mass. 
    
   Just as for S-V, Magneticum simulation overpredicts the stellar mass of the central dominant and the large offset BGGs  in observations by $\sim0.4\;dex$.
    \item  For S-III, we determine a normal distribution for the stellar mass of central dominant BGGs when compared with S-IV. It also appears that the peak of this distribution becomes more flat and its position tends to move to lower values, indicating that the fraction of BGGs at the left hand side of the distribution have been increased compared to that for S-IV.  This can be explained by the in-fall of massive galaxies to the groups or group mergers at this epoch. For S-III, the stellar mass distributions for the central BGGs and the large offset BGGs is approximately consistent. The G11 and H15 give good predictions for the position of the peak of the stellar mass distribution for the central BGGs. However, H15 underpredicts the observed mean stellar mass for the large offset BGGs by $0.3\;dex$. 
    
    Just as for S-V and S-IV, the Magneticum simulation over-predicts the stellar mass of the central dominant and the large offset BGGs  in observations by about $\sim0.4\;dex$.
     \item For S-II, we find a significant evolution in the shape of the stellar mass distribution for both the central dominant BGGs and the large offset BGGs. A second peak appears in the stellar mass distribution for central galaxies at  $log(M_\star/M_\odot)=10.5$, while the enhancements in the $log(M_\star/M_\odot)$ distribution for large offset BGGs at lower masses disappear and it roughly becomes as a normal distribution. We find that the observed mean stellar mass of the central dominant BGGs is higher than that of BGGs with a large offset of $0.5\;dex$. In the H15 prediction, this deviation is $0.25\;dex$. H15 well predicts  the mean stellar mass (the position of the centre of the peak) for BGGs with large offset. Both  H15 and G11 models under-predict the mean stellar mass of dominant central BGGs for S-II by $\sim0.4\;dex$. 
     
     It appears that the predictions by the Magneticum simulation becomes close to observation compared to those of the high-z susamples. However, there is still $\sim0.25\;dex$ differences between the stellar mass of BGGs in the observations and this simulation. 
   \item For S-I, the shape of the stellar mass distribution is similar for both  of the central dominant BGGs and BGGs with large offsets, however that of the centrals spans a wide dynamic mass range. Both distributions represent deviation from the normal distribution on the left-hand tail at lower stellar masses. The sign of the presence of the second peak in this side of the distribution is evident. The height of the peak and its position are consistent for both central BGGs and BGGs with large offsets. We observe similar trends in model predictions, however models under-predict the observed mean stellar mass by $0.5\;dex$. 

The Magneticum simulation predicts the stellar mass distribution of both the central and  the large offset BGGs for S-I remarkably well. 

\item On the right-hand side of the bottom panel of Fig. \ref{sm}, we show the stellar mass of BGGs with low offsets (black points) versus the redshift in observations. We also apply the KDE method and measure the corresponding density as shown with the shaded red contours (area). We determine the density for the central BGGs from the H15 model. The purple contours illustrate this density. We mention that G11 also predicts in the same way as the H15 model. We find that these models overall under-predict the stellar mass of BGGs in observations. The green contours presents the data for the central dominant BGGs in the Magneticum simulation. 
\item According to our observations, we argue that BGG is not always at rest at the centre of potential well of a virialised halo. We conclude that the BGG offset from X-ray centre of a group is as an important observable and parameter which is suggestive that it should be taken into account in galaxy formation model based on the CGP. The offset assumption can  bring  those model predictions more closer to the observed features of BGGs.
\end{enumerate}

In summary, observations and models indicate that the central dominant BGGs are generally more massive than the BGGs with large offsets. We find systematic differences between model predictions and observations which could have many reasons. The simulations in general should capture
this dynamical effect well, but the predicted x-ray emission (and 
therefore the definition of the centroid)
could suffer from too simplistic assumptions going in here. Obviously, it 
could be also the action of the AGN
which seems not to have been captured very well in the simulations. In addition, the
contamination of AGN or the contribution of ISM or metal lines in the 
colder phase could lead to more emission
from the BGG in reality than predicted by the simplistic approach in the 
simulations, and therefore change the
offset calculation. In observations, the samples
are flux limited, so observations could be 
biased to cool-core systems at high redshifts.
\begin{figure}
\includegraphics[width=0.49\textwidth]{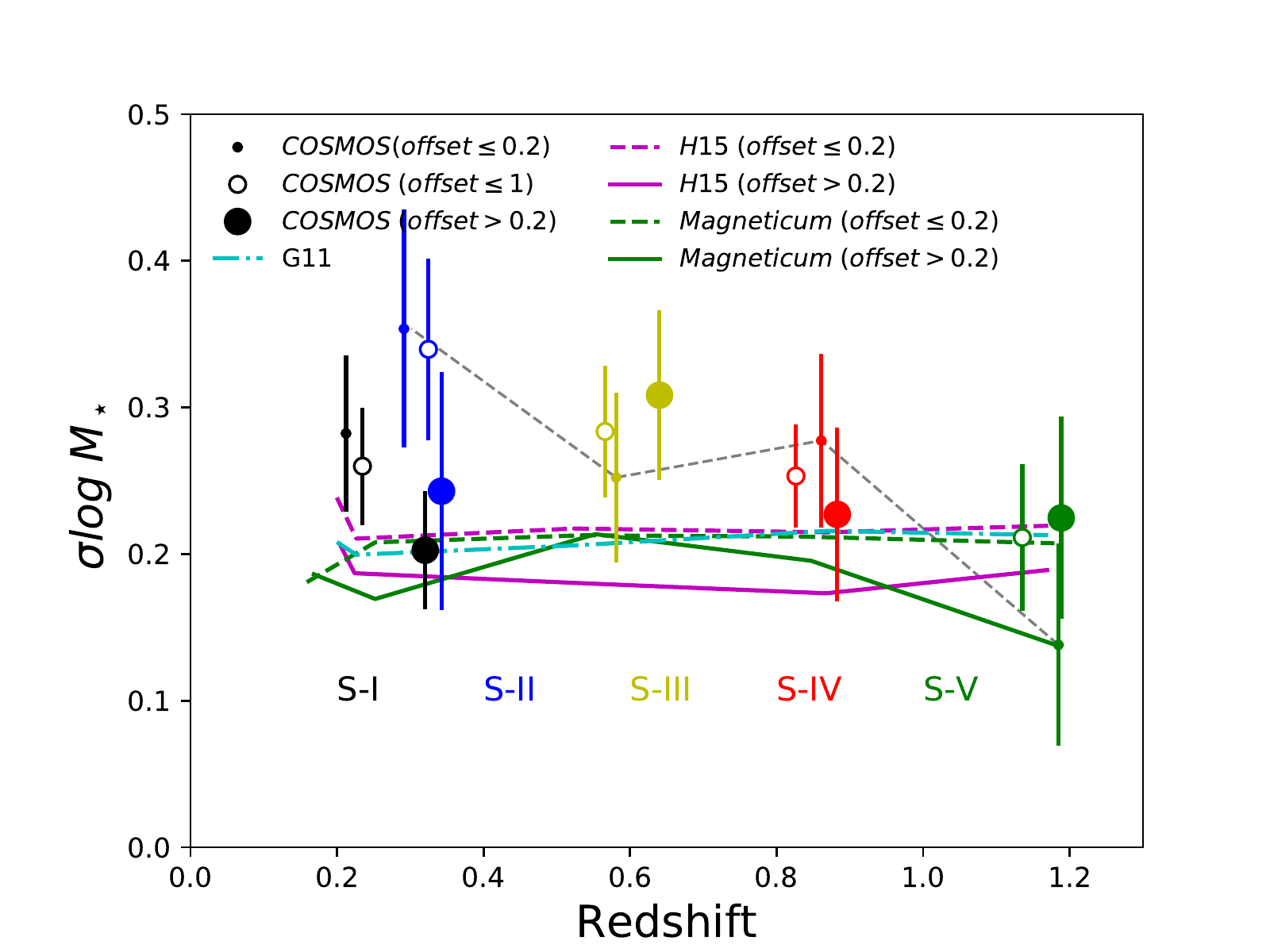}
\includegraphics[width=0.49\textwidth]{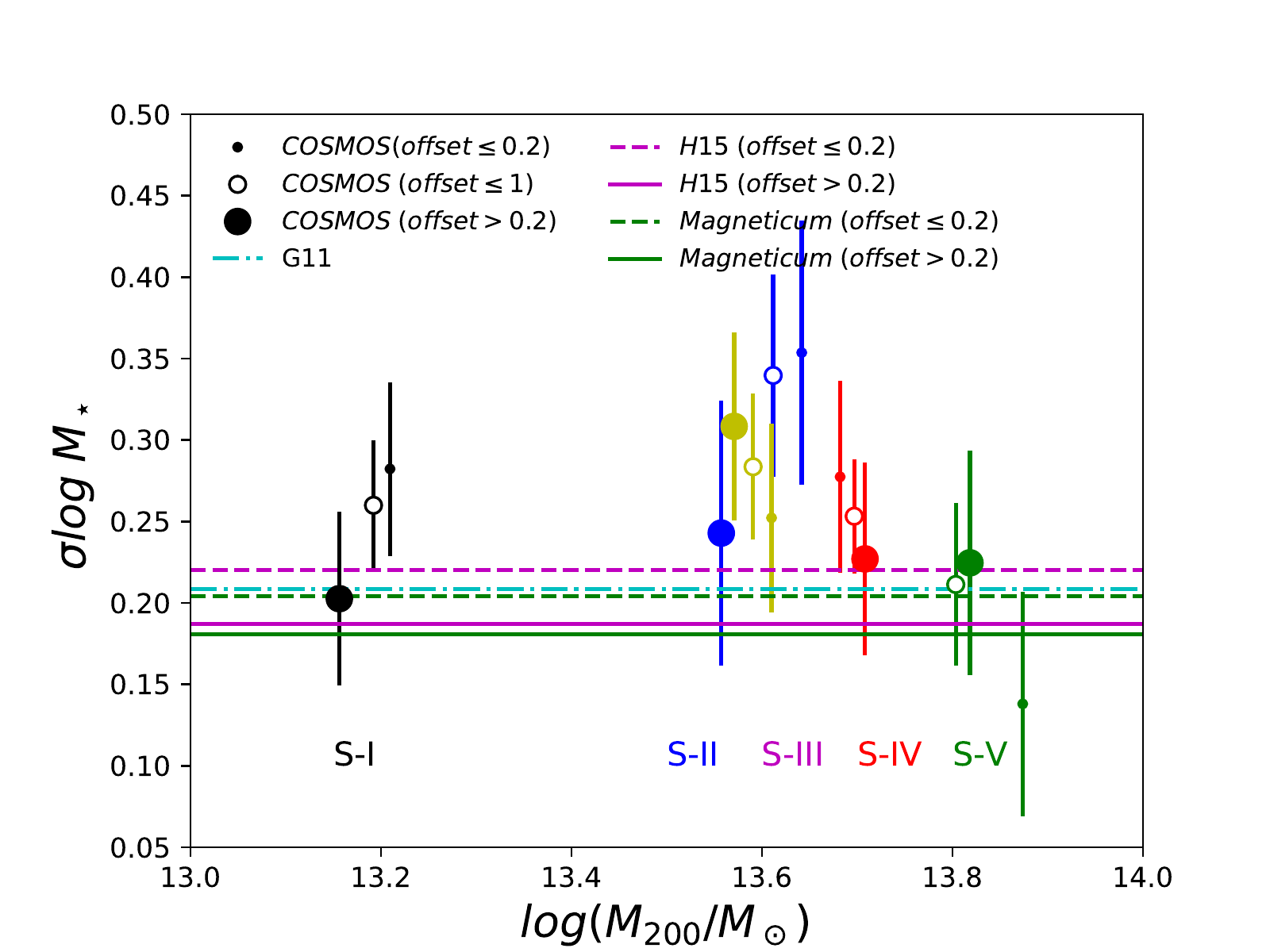}
\caption[]{ The scatter in the stellar mass of BGGs ($ \sigma _{log\;M_\star}$) as a function of redshift and halo mass. $ \sigma _{log\;M_\star}$ for the full sample of BGGs, the central BGGs (offset=$\dfrac{r [deg]}{R_{200}[deg]}\leq 0.2$) , and the large offset BGGs (offset$>0.2$) in observations are plotted as the open circles, filled small circles, and filled large circles, respectively. Results for S-I to S-V subsamples are shown with black, blue, yellow, red, and green symbols, respectively. The horizontal dashed and solid magenta lines and dash-dotted cyan line present constant scatter in the stellar mass of central dominant BGGs and BGG with large offsets from H15 model and the central BGGs (type=0 galaxies) from the SAMs of G11, respectively. The solid and dashed green lines shows the lognormal scatter in the stellar mass of the central dominant and the large offset BGGs in the Magneticum simulation.} \label{sigma}
\end{figure}	
\subsection {THE LOG-NORMAL SCATTER IN THE STELLAR MASS OF BGGS AT FIXED HALO MASS AND REDSHIFT}  \label{SHM}

The stellar mass of central galaxies  exhibits a tight relation with 
the halo mass  of hosting haloes. This relation is as an important observable and  constraining this relation is a key way to examine model predictions, recognizing the role of different physical mechanisms (e.g. star formation and feedback from stellar evolution and AGN activity) in the formation of BGGs
\citep{Yang12,Behroozi10,Behroozi13,Moster13,Coupon15}. Both observations and models illustrate that there exists a scatter in the stellar mass of central galaxies at fixed halo mass. This is one of the main source of difference between results, and the origin of this scatter is still an unresolved problem. In \citep{gozaliasl2018brightest}, we showed that the scatter in the stellar mass of BGGs is higher than the prediction by the H15 SAM and the study based on abundance matching and HOD methods by \cite[e.g.,][]{Coupon15}. The observed scatter increases with redshift from $\sigma_{log\; M_\star}\sim0.3$ at $z\sim0.2 $ to 0.5 at $ z\sim1.0 $. Our measurements show a remarkable agreement with recent a study by 
 \cite{Chiu2016b} who measured the $ M_\star-M_{200} $  scaling relation for 46 
 X-ray  
 groups detected in the XMM-Newton-Blanco Cosmology Survey (XMM-BCS) with a median halo mass of $ 8 \times 10^{13} M_\odot $) at a median 
 redshift of z=0.47, finding a scatter of $ \sigma_{log M_\star|M_{500}}=0.36^{+0.07}_{-0.06}$. In this study, we used data from different surveys and we measured the scatter for all BGGs as a whole.

 Using this revised  data of X-ray galaxy groups and BGGs in the COSMOS field, we recalculate the log-normal scatter in the stellar mass for the full sample of BGGs, BGGs with low and large offsets at fixed redshift and halo mass. We investigate whether the offset between the position of the BGGs and the X-ray centres of their host haloes might have an impact on scattering the BGG mass.

 Figure \ref{sigma} shows the log-normal scatter in stellar mass of BGGs as a function 
 of redshift (upper panel) and halo mass $ M_{200}$ (lower panel). We note that the scatter corresponds to the standard deviation of  
 $log(M_\star/M_\odot)$ at a given redshift/halo mass range. The results for different subsamples are shown with  different colours. To distinguish data points of BGGs with different 
 offsets values, we shift data along the x-axis by $\pm0.03 $. We compare our results from the observations with predictions from the G11 and H15 SAMs. We determine scatter in the stellar mass of the central BGGs with offset$\leq0.2$ (dashed magenta line) and BGGs with large offset$>0.2$ (solid magenta line) in the H15 model and the central BGGs (type=0) in the G11 model (dashed cyan line). The solid and dashed green lines illustrate the scatter in the stellar mass of the central and offset  BGGs in the Magneticum simulation.
In the upper panel of Fig. \ref{sigma}, we find that $\sigma_{log\;M\star}$ for all types of BGGs in terms of offset are generally consistent and the major differences in $\sigma_{log\;M\star}$ between the large offset and the low offset BGGs are seen for S-IV, S-II, and S-I by approximately 0.1 dex which is not significant within errors.  $\sigma_{log\;M\star}$ for S-II to S-IV remain constant around $\sigma_{log\;M\star}\sim0.30\pm0.07$ dex  at $z<1$. While $\sigma_{log\;M\star}$
 for S-V drops to $\sigma_{log\;M\star}\sim 0.1\pm0.06$, indicating a redshift evolution between S-V and S-IV (from $z\sim1.3$ to $z\sim0.7$ ). Interestingly, when the current measurement of $\sigma_{log\;M\star}$ for S-V is compared with our previous measurement ( $\sigma_{log\;M\star}\sim 0.5\pm0.09$ ) in \cite{gozaliasl2018brightest}, we find that the current estimate of $\sigma_{log\;M\star}$ is much lower than the former estimate. This indicates that the quality of the galaxy stellar mass estimates in COSMOS is much better than  the data of galaxies in XMM-LSS and AEGIS field, which we used in \cite{gozaliasl2018brightest}. For S-I, we also have an improvement in our measurement compared to that in \cite{gozaliasl2018brightest} and find that the current estimate $\sigma_{log\;M\star}$ for S-I is in agreement with other subsamples of BGGs (e.g., S-II) at $z<1$.

 For the central dominant and the large offset BGGs selected from the H15 model, we estimate that $\sigma_{log\;M\star}$ remain constant with both redshift and halo mass at  $\sigma_{log\;M\star}=0.220$  and $0.187\;dex$, respectively. It appears that the scatter in the stellar mass of BGGs with large offset in H15 model is $\sim0.04\;dex$ less than that of the central BGGs in this model. The scatter in the stellar mass of central BGG in the G11 prediction is  $\sigma_{log\;M\star}=0.208\;dex$. Models are found to be in a good agreement with the data in observations within errors.
In the lower panel of Fig. \ref{sigma}, we find no significant dependence of  $\sigma_{log\;M\star}$ to halo mass, however, noting that the halo mass range of groups used in this analysis is too small. 
 The scatters of the stellar mass of the central dominant (dashed green line) and the offset (solid green line) of BGGs in the Magneticum simulation are consistent observations within the errors and predictions from the H15 and G11 models. In the Magneticum simulation, we find that the scatter in the stellar mass of the large offset BGGs increase slightly with redshift which is in a good agreement with observations. 
\section{Summary and conclusions}
We present the revised catalogue of 247 X-ray groups of galaxies in the 2 square degree COSMOS field with $M_{200c}=8\times10^{12}-3\times10^{14} \;M_{\odot}$ at redshift range of $ 0.08\leq z < 1.3$. The main revisions are on the group X-ray centre using the combined data of the XMM-Newton and Chandra and the redshift based on the COSMOS2015 photometric redshifts catalogue \citep{laigle2016cosmos2015} and the COSMOS spectroscopic redshifts catalogue \citep{hasinger2018}. We select the brightest group 
galaxies from our X-ray galaxy groups and define five 
subsamples (S-I to S-V) considering the halo mass and redshift of hosting groups such that four out of five have the same halo 
mass range. This definition allows us to investigate the stellar mass distribution of BGGs within  
haloes of similar masses, but at different redshifts. We study differences in the stellar mass distribution between the central dominant BGGs and BGGs with a large offset from the X-ray peak. The BGG offset is defined as the ratio of the separation between the position of this galaxy and the group X-ray centre to the group $R_{200}$ radius. BGGs in each subsample are classified into three types based on the offset: the central dominant BGGs (offset $\;\leq 0.2$), large offset BGGs (offset offset$>0.2$) , and  full sample of BGGs (offset $\leq 1$).  We determine the log-normal scatter in the stellar mass of BGGs. We interpret our results with predictions 
from two SAMs of H15 and G11 implemented based on the Millennium simulation and the results from the hydrodynamical simulation of Magneticum. The summary of our findings is as follows:
\begin{enumerate}
\item We inspect the BGG  selection within different radii from the X-ray centre of haloes ($R_{200}, R_{500}, 0.5R_{500}$) and find that the best aperture for the BGG selection for groups with $M_{200}\sim10^{13-14}M_\odot$ is $R_{200}$. By decreasing the aperture from  $R_{200} $ to $ 0.5R_{500}$, the BGG stellar mass tends to skew towards low masses and the probability of missing true BGG increases from 40\% to 50\%. When selecting BGGs within $0.5R_{500}$, consequently, for a number of groups, centrally located low mass satellite galaxies  with $M_*\sim10^{8-9} M_\odot$ are selected as BGGs while there are more massive galaxies at $100-300\; kpc$. Although, the stellar mass distributions of the true (bona fide) BGGs selected within the mentioned apertures are similar above $M_*\sim10^{10.5} M_\odot$. 
\item  We find that the BGG offset decreases by a factor of three from $z=1.3$ to the present day. We visually inspect the multiband images of groups having BGGs with large offsets and find that they generally include two massive and luminous galaxies. We conclude that these bright group members finally merge into one with time and the newly formed BGG  becomes closer to the host group X-ray centre.
\item We measure the r-band  magnitude gap between the first and the second brightest group galaxies within $R_{200}$ and investigate its relation with the offset of the first brightest group galaxy from the X-ray centre. We find that  the offset decreases as a function of increasing  magnitude gap with no considerable redshift dependent trend.
\item We classified our sample of groups into three redshift bins, $z=0.0-0.5, z=0.5-1.0$, and $z=1.0-1.5$ and selected clean groups in which we are able to define their X-ray centres. We found that the BGG offset from the group's X-ray centre decreases as a function of increasing group total mass ($M_{200}$) and the slope of the relation increases with increasing redshift. We show that the offset is not an effect driven by lower SNR and it shows no dependence on the X-ray flux and flux significance. 
	\item We applied the normality test and find that the $ log(M_\star/M_\odot)$ distributions for the full sample of BGGs for S-I, S-III, and SIV  deviates a little from a normal distribution. This deviation in the shape of the stellar mass distribution is due to the deviation of the shape of the stellar mass distribution of BGGs with large offset, in particular, at $z>0.4$. However, at $z<0.4$, the  distribution of BGGs with low-offset for S-II leads the stellar mass distribution of the full sample of BGGs to deviate from a Gaussian distribution. We observe a second peak in the stellar mass distribution of the central dominant BGGs for S-II at $z<0.4$.
  \item By comparing the  $log(M_\star/M_\odot)$ distribution between BGGs with low-offset with that of BGGs with large-offset, we conclude that  the central BGGs are not evolving in the same fashion as BGGs with large offsets. Clearly, the differences between stellar mass distributions of BGGs with small and large offsets suggest that the offset is an important observable which must be taken into account in modelling BGGs/BCGs and  hosting haloes as well.  We believe that there are several astrophysical phenomena such major merger, group/halo merger and in-falling  massive galaxies into a system, all can lead to a large offset among the BGG position and the group X-ray centre.

	\item Using our unique sample of BGGs, we determine a constant log-normal scatter in the stellar mass of BGGs,  $\sigma_{log\;M\star}\sim0.30\pm0.07$ dex, at $z<1.0$ with no significant dependence on the BGG offset from the group X-ray centre. This scatter interestingly decreases to $\sigma_{log\;M\star}\sim0.10\pm0.06$ at $z=1.0-1.3$ for our S-V subsample, indicating a little redshift evolution from z=1.3 to z=0.17. The $\sigma_{log\;M\star}$ which we measure here is up to 0.15 dex less than that we estimate in our recent measurement \citep{gozaliasl2018brightest} in the same redshift and halo mass ranges. We conclude that the high quality multi-bands data of COSMOS effectively decrease bias in the stellar mass measurement and mixing low and high quality data from different surveys may lead to a large bias and scatter in the  $\sigma_{log\;M\star}$ measurement , even, if a similar method is used for estimating the stellar mass of galaxies from different surveys. Multi-band observations and a precise redshift determination of galaxies are vital in measuring their stellar properties.  
    
   \item  We find that the scatter in the stellar mass of BGGs does not depend significantly on the BGG offsets  from the group X-ray centres. 
   \item By comparing our results with those from two SAMs of H15 and G11 and the hydrodynamical simulation of Magneticum, we conclude that models have generally captured the observed trends. Notably, we find that the mean stellar mass of the central dominant BGGs is higher than that of the large offset BGGs in a good agreement with model predictions. However, there is still a systematic differences between the predictions from simulations and observations which can arise for several reasons. For instance, the action of the AGN seems have not been captured very well in the simulations. In addition, the
contamination of AGN or the contribution of ISM or metal lines in the 
colder phase could lead to more emission
from the BGG in reality than predicted by the simplistic approach in the 
simulations, and therefore change the
offset calculation. In the observations, the samples
are flux limited, so observations could be 
biased to cool-core systems at high redshifts. 
\end{enumerate}
\section{Acknowledgements}
This work has been supported by grants from the Finnish Academy of Science to the University of Helsinki and the Euclid project, decision numbers 266918 and 1295113. G. G. wishes to thank Mr. Donald Smart for very useful comments. S.T. acknowledge support from the ERC Consolidator Grant funding scheme (project Context, grant No. 648179). The Cosmic Dawn Center is funded by the Danish National Research Foundation. We used data of the Millennium simulation from the web application providing on-line access, constructed as activities of the German Astrophysics Virtual Observatory.
\bibliographystyle{mn2e}
\bibliography{ggcite}

\textbf{\\$^{*}$The rest of affiliations from the first page:\\}
  $^{4}$ Max Planck-Institute for Extraterrestrial Physics, P.O. Box 1312, Giessenbachstr. 1., D-85741 Garching, Germany\\
  $^5$ School of Astronomy, Institute for Research in Fundamental Sciences (IPM), Tehran, Iran\\
$^6$ Argelander-Institut f\"{u}r Astronomie, Universit\"{a}t Bonn, Auf dem Hügel 71, D-53121 Bonn, Germany\\ 
 $^7$ National Astronomical Observatory of Japan, 2-21-1 Osawa, Mitaka, Tokyo 181-8588, Japan\\
 $^8$ Aix Marseille Universit\'{e}, CNRS, Laboratoire \v{d} Astrophysique de Marseille, UMR 7326, F-13388 Marseille, France\\ 
 $^{9}$ CNRS, UMR 7095 \& UPMC, Institut \v{d}Astrophysique de Paris, 98bis boulevard Arago, 75014 Paris, France\\
$^{10}$ Physics Department, University of Miami, Knight Physics Building, Coral Gables, FL  33124, USA\\
$^{12}$ Universit\"ats-Sternwarte M\"unchen, Scheinerstr.\ 1, D-81679 M\"unchen, Germany\\
$^{13}$ Max Planck Institut for Astrophysics, D-85748 Garching, Germany\\
$^{14}$ CEA Saclay, Laboratoire AIM-CNRS-Universit\'{e} Paris Diderot, Irfu/SAp, Orme des Merisiers, F-91191, Gif-sur-Yvette, France\\
$^{15}$ European Space Astronomy Centre (ESA/ESAC), Director of Science, E-28691 Villanueva de la Ca\~{n}ada, Madrid, Spain\\
$^{16}$ IPAC, Mail Code 314-6, California Institute of Technology, 1200 East California Boulevard, Pasadena, CA 91125, USA\\
$^{17}$ Cosmic Dawn centre (DAWN), Niels Bohr Institute, University of Copenhagen, Juliane Maries vej 30, DK-2100 Copenhagen, Denmark\\
$^{18}$ Sub-department of Astrophysics, University of Oxford, Keble Road, Oxford OX1 3RH, UK\\
$^{19}$ California Institute of Technology, 1200 E. California Boulevard, Pasadena, CA, 91125, USA\\
$^{20}$ Harvard-Smithsonian centre for Astrophysics, 60 Garden Street, Cambridge, MA 02138, USA\\
$^{21}$ Department of Physics \& Astronomy, University of Hawaii at Hilo, 200 W. Kawili Street, Hilo, HI 96720, USA\\
$^{22}$Department of Physics and Astronomy, University of Waterloo, 200 University Avenue West, Waterloo, Ontario, Canada N2L 3G1, CAN\\
$^{23}$ Institute for Astronomy, University of Hawaii at Manoa, Honolulu, HI 96822, USA\\
$^{24}$ University of Paris Denis Diderot, University of Paris Sorbonne Cit\'{e} (PSC), 75205 Paris Cedex 13, France\\
$^{25}$ Sorbonne Universit\'{e}, Observatoire de Paris, Universit\'{e} PSL, CNRS, LERMA, F-75014, Paris, France\\
$^{26}$ Jet Propulsion Laboratory, Cahill centre for Astronomy $\&$ Astrophysics, California Institute of Technology, 4800 Oak Grove Drive, Pasadena, California, USA\\
$^{27}$ INAF-Osservatorio Astronomico di Brera, via Brera 28, I-20122 Milano,
Italy\\
$^{28}$ Institute for Astronomy, ETH Zurich, CH-8093 Zurich, Switzerland\\
$^{29}$ Cosmic Dawn Center (DAWN), Niels Bohr Institute, University of Copenhagen, DK-2100 Copenhagen \o; DTU-Space, Technical University of Denmark, DK-2800 Kgs. Lyngby\\
\begin{landscape}
\begin{table}
 \caption{ The revised catalogue of X-ray galaxy groups in 2 square degree of the COSMOS field which previously presented by Finoguenov et al. (2007)
  and George et al. (2011). The full catalogue is available in online edition.}

\begin{tabular}{|l|l|l|l|l|l|l|l|l|l|l|l|l|}

\hline
ID-COSMOS  &Ra &Dec& redshift($z$) & flag & $M_{200c}$ &  $L_{X}$& $R_{200}$&$kT$&$Flux$&$Flux significance$& redshift type\\
 &$[J2000]$  &$[J2000]$  &  &  & $[\times 10^{12}\;M_\odot]$ &  $[\times10^{42}\;erg\; s^{-1}$] & $[deg]$ &$[keV]$&$[\times10^{-15}\;erg\;cm^{-2}\; s^{-1}]$& & \\
\hline
  10006 & 150.29951 & 1.55343 & 0.360     & 1       & 74.57 $\pm$4.17   & 11.53$\pm$1.02  & 0.0425   & 1.41$\pm$0.05 & 15.96 $\pm$1.42   & 11.26    & spec\\
  30010 & 150.75047 & 1.53103 & 0.669    & 1       & 134.42$\pm$23.14  & 43.85$\pm$12.35 & 0.0332   & 2.32$\pm$0.27 & 14.44 $\pm$4.07   & 3.55   & spec\\
  20011 & 150.19728 & 1.65895 & 0.220     & 1       & 170.01$\pm$1.73   & 34.98$\pm$0.56  & 0.0838   & 2.33$\pm$0.02 & 159.56$\pm$2.55   & 62.58   & spec\\
  20017 & 149.96375 & 1.68022 & 0.373    & 1       & 95.64 $\pm$4.18   & 17.21$\pm$1.19  & 0.0453   & 1.66$\pm$0.05 & 22.69 $\pm$1.57   & 14.46    & spec\\
  30018 & 149.76518 & 1.62355 & 0.371    & 2       & 33.44 $\pm$6.68   & 3.33 $\pm$1.10   & 0.032    & 0.87$\pm$0.10  & 4.33  $\pm$1.43   & 3.04     & spec\\
  20020 & 150.32584 & 1.60510  & 0.227    & 2       & 21.19 $\pm$2.62   & 1.35 $\pm$0.27  & 0.0419   & 0.65$\pm$0.04 & 6.16  $\pm$1.23   & 5.01     & spec\\
  20023 & 150.38068 & 1.67576 & 0.701    & 1       & 51.63 $\pm$5.79   & 10.30 $\pm$1.86  & 0.0233   & 1.25$\pm$0.09 & 2.63  $\pm$0.48   & 5.53   & spec\\
  20024 & 150.29169 & 1.68935 & 0.527    & 1       & 49.85 $\pm$4.31   & 7.65 $\pm$1.06  & 0.0281   & 1.16$\pm$0.06 & 4.10  $\pm$0.57   & 7.22     & spec\\
  20025 & 149.85402 & 1.77023 & 0.124    & 1       & 44.33 $\pm$0.88   & 3.84 $\pm$0.12  & 0.0884   & 0.96$\pm$0.01 & 63.14 $\pm$1.98   & 31.94    & spec\\
  20029 & 150.18167 & 1.77086 & 0.344    & 1       & 64.47 $\pm$2.41   & 8.96 $\pm$0.53  & 0.0422   & 1.28$\pm$0.03 & 14.09 $\pm$0.83   & 16.91   & spec\\
  10031 & 150.49237 & 1.79425 & 0.958    & 2       & 66.28 $\pm$8.04   & 21.82$\pm$4.27  & 0.0207   & 1.61$\pm$0.13 & 2.51  $\pm$0.49   & 5.11     & spec\\
  20035 & 150.21020  & 1.82014 & 0.529    & 1       & 47.09 $\pm$4.38   & 7.02 $\pm$1.05  & 0.0275   & 1.12$\pm$0.06 & 3.69  $\pm$0.55   & 6.71  & spec\\
  20039 & 149.82126 & 1.82543 & 0.531    & 1       & 39.97 $\pm$4.58   & 5.44 $\pm$1.01  & 0.0260    & 1.01$\pm$0.07 & 2.83  $\pm$0.52   & 5.42    & spec\\
  10040 & 150.41250  & 1.84888 & 0.973    & 1       & 82.31 $\pm$6.07   & 31.08$\pm$3.65  & 0.0221   & 1.86$\pm$0.09 & 3.67  $\pm$0.43   & 8.51   & spec\\
  10052 & 150.44759 & 1.88319 & 0.672    & 1       & 41.22 $\pm$5.44   & 6.93 $\pm$1.48  & 0.0223   & 1.08$\pm$0.09 & 1.93  $\pm$0.41   & 4.68   & spec\\
  10054 & 150.58664 & 1.93761 & 0.310     & 2       & 40.07 $\pm$3.36   & 4.08 $\pm$0.55  & 0.0392   & 0.95$\pm$0.05 & 8.32  $\pm$1.12   & 7.46   & spec\\
  20058 & 150.12393 & 1.91144 & 0.736    & 1       & 35.10  $\pm$7.19   & 5.91 $\pm$2.00   & 0.0199   & 1.00 $\pm$0.12 & 1.24  $\pm$0.42   & 2.96   & spec\\
  10064 & 150.19821 & 1.98506 & 0.440     & 1       & 32.05 $\pm$3.24   & 3.41 $\pm$0.55  & 0.0278   & 0.87$\pm$0.05 & 2.88  $\pm$0.47   & 6.16   & spec\\
  20065 & 149.89253 & 1.94459 & 0.773    & 1       & 28.96 $\pm$6.99   & 4.61 $\pm$1.85  & 0.0181   & 0.91$\pm$0.12 & 0.78  $\pm$0.31   & 2.49    & spec\\
  20067 & 149.74410  & 1.95229 & 0.483    & 5       & 26.80  $\pm$5.06   & 2.69 $\pm$0.83  & 0.0248   & 0.79$\pm$0.08 & 1.86  $\pm$0.58   & 3.22   & spec\\
  20068 & 149.99960  & 1.97328 & 0.600      & 1       & 32.52 $\pm$5.31   & 4.34 $\pm$1.16  & 0.0223   & 0.92$\pm$0.09 & 1.59  $\pm$0.42   & 3.75  & spec\\
  20069 & 150.41965 & 1.97382 & 0.863    & 1       & 29.82 $\pm$9.86   & 5.48 $\pm$3.08  & 0.0170    & 0.95$\pm$0.18 & 0.64  $\pm$0.36   & 1.78& spec\\
  20071 & 150.36566 & 2.00252 & 0.828    & 1       & 53.59 $\pm$6.25   & 13.05$\pm$2.45  & 0.0212   & 1.34$\pm$0.10  & 2.13  $\pm$0.40    & 5.32   & spec\\
  10073 & 150.09186 & 1.99908 & 0.221    & 1       & 18.52 $\pm$1.64   & 1.10  $\pm$0.16  & 0.0397   & 0.61$\pm$0.03 & 4.86  $\pm$0.69   & 7.06   & spec\\
    10041  &     149.9323  &  1.83045  &  0.530   &    2   &   24.82$\pm$5.03  &     2.582$\pm$ 0.86  &  0.0222   &    0.77$\pm$  0.08   &  1.28 $\pm$   0.43 &     2.99  &        spec* \\  
   10105  &     150.38295  & 2.10278  &  1.163  &    1   &   61.21 $\pm$    10.48  &    25.61 $\pm$    7.17  &  0.0179  &   1.64 $\pm$   0.18   &  1.64 $\pm$     0.46 &     3.57      &    phot\\

        \hline
 		\end{tabular}\label{catalogue}
        \end{table}
        \footnotesize{Column 1 lists identification ID. Columns 2 and 3 present right ascension and declination of the X-ray peak (centre) in equinox $J2000.0$. Column 4 and 5 present the redshift and group's identification flags.  Column 6 and 7 list $M_{200}$ with $\pm 1 \sigma$ error in $[\times 10^{12} M\sun]$ and the X-ray luminosity ($L_X$) with $\pm 1 \sigma$ error in [$\times 10^{42}erg\;s^{-1}$] within $r_{500}$. Column 8 reports group $R_{200}$ in degree. Columns 9 and 10 report the IGM temperature ($kT$) with corresponding $\pm1\sigma$ error in $keV$ and the X-ray flux in $0.5-2\; keV$ band within $r_{500}$ in $[\times 10^{-15}\; erg\; cm^{-2}\; s^{-1}]$ with $\pm1\sigma$ errors. Column 11 provides the significance of the X-ray flux estimate. Column 12 presents two types of the assigned redshift to the groups:$`spec'-z$ or $`phot'-z$. The redshift of groups with the $`spec*'-z$ are measured using their $spec-z$ members within $R_{200}$. The redshift of the rest of groups with "$spec-z$" are determined using their $spec-z$ members within $R_{500}$.  We note that the errors are presented in separate columns in the electronic version of the catalogue.}\\
\end{landscape}

\end{document}